\documentclass[ALICE,manyauthors]{cernphprep}

\usepackage[comma,square,numbers,sort&compress]{natbib}
\usepackage{hyperref}
\usepackage{lineno}
\usepackage{pennames}
\usepackage{hepnames}
\newcommand{\rb}[1]   {\mbox{\textrm{\scriptsize #1}}}
\newcommand{\rbt}[1]  {\mbox{\textrm{\tiny #1}}}
\newcommand{\fb}      {\ensuremath{f_{\rbt{b}}}}
\newcommand{\fbprime} {\ensuremath{f_{\rbt{b}}^{\prime}}}
\newcommand{\fsig}    {\ensuremath{f_{\rbt{Sig}}}}
\newcommand{\ffb}     {\ensuremath{F_{\rbt{Bkg}}}}
\newcommand{\ffs}     {\ensuremath{F_{\rbt{Sig}}}}

\newcommand{\gevc}    {\mbox{GeV$/c$}}
\newcommand{\gevcc}   {\mbox{GeV$/c^2$}}

\newcommand{\lumint}  {\ensuremath{{\cal L}_{\rb{int}}}}
\newcommand{\acceff}  {\ensuremath{{A \times \epsilon}}}
\newcommand{\sqrts}   {\ensuremath{\sqrt{s}}}
\newcommand{\sqrtsnn} {\ensuremath{\sqrt{s_{_{\rbt{NN}}}}}}
\newcommand{\pt}{$p_{\rm T}$}
\newcommand{\Pt}{p_{\rm T}}

\newcommand{\dd}{\mathrm{d}}

\begin{document}%

\begin{titlepage}
\PHyear{2018}
\PHnumber{010}      
\PHdate{30 January}  
%

\title{%
	Prompt and non-prompt \PJpsi production and nuclear modification at mid-rapidity in 
	p-Pb collisions at ${\bf \sqrt{{\it s}_{\text{NN}}}= 5.02}$~TeV
	}
\ShortTitle{Prompt and non-prompt \PJpsi production in p-Pb collisions at $\sqrt{s_{\rm NN}}=5.02$ TeV}   

\Collaboration{ALICE Collaboration\thanks{See Appendix~\ref{app:collab} for the list of collaboration members}}
\ShortAuthor{ALICE Collaboration} 

\begin{abstract}
	A measurement of beauty hadron production at mid-rapidity in proton-lead 
	collisions at a nucleon-nucleon centre-of-mass energy $\sqrt{s_{\rm NN}}=5.02$ TeV is presented. 
	The semi-inclusive decay channel  
	of beauty hadrons into \PJpsi  
	is considered, where the \PJpsi mesons are reconstructed in the 
	dielectron decay 
	channel at mid-rapidity down to transverse momenta of 1.3~GeV/$c$. The $\Pbottom\APbottom$ production cross section 
	at mid-rapidity, $\dd\sigma_{\Pbeauty\APbeauty}/\dd y$,  and the total cross section extrapolated over full phase 
	space, $\sigma_{\rm b\bar{b}}$, are obtained.  
	This measurement  is combined with results on inclusive \PJpsi production to determine the prompt \PJpsi cross sections. 
	The results in p-Pb collisions are then scaled to expectations from pp collisions at the same centre-of-mass energy to derive 
	the nuclear modification factor $R_{\rm pPb}$,  
	and compared to models to study possible nuclear modifications of the production 
	induced by cold nuclear matter effects.  
	$R_{\rm pPb}$ is found to be smaller than unity at low \pt\
	for both \PJpsi coming from beauty hadron decays and prompt \PJpsi.    
\end{abstract}
\end{titlepage}
\setcounter{page}{2}

\section{Introduction}

In high-energy hadronic collisions the production of beauty-flavoured hadrons, referred to as b-hadrons ($\rm h_b$) in the 
following, represents a challenging testing ground for models based on quantum chromodynamics (QCD). 

In proton-proton (pp) collisions the production cross sections can be computed with a factorisation approach~\cite{Factorisation1,Factorisation2}, 
as a convolution of the parton distribution functions (PDFs) 
of the incoming protons, the partonic hard-scattering cross sections, and the fragmentation functions.  

In ultra-relativistic heavy-ion collisions, where the formation of a high-density 
colour-deconfined medium, the Quark-Gluon Plasma (QGP), is expected~\cite{Cabibbo,Shuryak}, heavy quarks  
are considered as prime probes of the properties of the medium created in the collision. Indeed, they are produced in scattering processes 
with large momentum transfer in the first stage of the collision 
and traverse the medium interacting with its constituents, 
thus experiencing its full evolution.  
Modifications in the production of b-hadrons 
with respect to expectations from  
an incoherent superposition of 
elementary pp collisions  
can reveal  
the properties of the medium.  
However, other effects, which are not related to the presence of a QGP, the so called Cold Nuclear
Matter (CNM) effects, can modify b-hadron production 
in heavy-ion collisions.   
In the initial state, the nuclear environment affects the free nucleon PDFs, which are modified 
depending on the parton fractional momentum $x_{\rm B}$, the four-momentum transfer squared and 
the atomic mass number A, as it was first 
observed by the European Muon Collaboration~\cite{EMC}. 
At the Large Hadron Collider (LHC) 
energies, the most relevant effects are 
parton-density shadowing or gluon saturation,  
which can be described using modified parton distribution functions in the nucleus~\cite{shadowing}
or using the Color Glass Condensate (CGC) effective theory~\cite{CGC,CGCreview}.
Partons can also lose energy in the early stages of the collision via initial-state
radiation, thus modifying the centre-of-mass energy of the partonic system~\cite{Vitev}, or experience transverse
momentum broadening due to multiple soft collisions before the $\Pbottom\APbottom$ pair is produced~\cite{Lev,Wang,Kopeliovich}. 

Measurements in proton-nucleus (p-A) collisions and their comparison to pp results provide a tool to constrain the CNM  
effects. 
To quantify these effects, the nuclear modification factor can be defined as  
the production cross section in p-A collisions ($\sigma_{\rm pA}$) divided by that in pp collisions ($\sigma_{\rm pp}$)
scaled by 
the atomic mass number A  
\begin{equation}
	R_{\rm pA}(y,p_{\rm T}) = 
    \frac{1}{\rm A}
	\frac{{\rm d}^2\sigma_{\rm pA} / {\rm d}y {\rm d}p_{\rm T}}
             {{\rm d}^2\sigma_{\rm pp} / {\rm d}y {\rm d}p_{\rm T}}  ,
\label{RpA}
\end{equation}
where $y$ is the rapidity of the measured hadron in the nucleon-nucleon centre-of-mass frame, and \pt\ its transverse momentum.   
In the absence of nuclear effects $R_{\rm pA}$ is expected to equal unity.  

Cross sections for beauty production in proton-nucleus collisions have been measured at fixed target experiments  
with beam energies 
of 800 and 
920~GeV~\cite{E789,E771,HERAB}, corresponding to nucleon-nucleon centre-of-mass energies up 
to $\sqrt{s_{\rm NN}}=41.6$~GeV. 
Measurements at the LHC in p-Pb collisions 
 are sensitive to a previously unexplored parton kinematic domain of the colliding nucleons, in particular to  
small values of the gluonic content of the nucleon $x_{\rm B}$. 
For instance, in the perturbative QCD leading order process $\Pgluon\Pgluon \rightarrow \Pbottom\APbottom$  
the  threshold production of a $\Pbottom\APbottom$ pair at $y=0$\ and $y=3$ in p-Pb collisions at $\sqrt{s_{\rm NN}}=5.02$~TeV  
is obtained, respectively, for $x_{\rm B}\approx 10^{-3}$ and $10^{-4}$~\cite{PPR2}. 
The LHCb experiment has measured beauty production  at backward and forward rapidity~\cite{LHCb,LHCb8TeV}, where
``forward'' and ``backward'' are defined relative to the direction of the proton, 
reporting 
$R_{\rm pPb}=0.83\pm 0.08$ at forward rapidity ($1.5<y<4$) and $  R_{\rm pPb}=0.98\pm 0.12$ at backward rapidity ($-5<y<-2.5$) 
in p-Pb collisions at $\sqrt{s_{\rm NN}}=5.02$~TeV.  
Results at mid-rapidity have been reported from the 
ATLAS and CMS experiments, 
based  on either exclusively reconstructed beauty mesons~\cite{CMSBmeson}, or semi-inclusive decays ${\rm h_b} \rightarrow \PJpsi + X $~\cite{ATLAS,CMS,ATLAS8TeV}  
or beauty jets~\cite{CMSjet}. These measurements however do not cover, at mid-rapidity, the low \pt\ region where the nuclear 
effects are expected to be 
the largest and the bulk of the total b-hadron production is concentrated.  
ALICE has measured beauty production in p-Pb collisions at $\sqrt{s_{\rm NN}}=5.02$~TeV through the semi-leptonic 
decay channel, ${\rm h_b} \rightarrow \Pe + X $, down to a  
transverse momentum of the decay electron of 1~GeV/$c$, finding $R_{\rm pPb}$ compatible with unity within 
large experimental uncertainties~\cite{ALICEele}.   

In this paper, the measurement of beauty production at mid-rapidity in p-Pb collisions at $\sqrt{s_{\rm NN}}=5.02$~TeV 
using the semi-inclusive channel ${\rm h_b} \rightarrow \PJpsi + X $\ is presented. 
The \PJpsi mesons are reconstructed in the dielectron decay channel, $\PJpsi \rightarrow \Pep\Pem$, down to 
\pt\ of 1.3 GeV/$c$ and for \PJpsi rapidity in the nucleon-nucleon centre-of-mass system within $-1.37 <y< 0.43$.  
The covered \pt\ range corresponds     
to about 80\% of the \pt-integrated cross section at mid-rapidity, $\dd \sigma / \dd y$, 
which allows to derive the \pt-integrated $\Pbeauty\APbeauty$ cross section $\dd\sigma_{\Pbeauty\APbeauty}/\dd y$ with 
extrapolation uncertainties of a few percent.  

ALICE already reported measurements of inclusive \PJpsi production at backward, mid- and forward rapidity 
in p-Pb collisions at $\sqrt{s_{\rm NN}}=5.02$~TeV down to $\Pt=0$~\cite{ALICEinclusive}.  
The production of the prompt \PJpsi meson in hadronic interactions represents another 
test for QCD-inspired models (for comprehensive reviews see, e.g.~\cite{BRAMBILLA,SAPOREGRAVIS}).  
The inclusive \PJpsi yield is composed of three contributions: prompt \PJpsi produced directly in the p-Pb 
collision, prompt \PJpsi produced indirectly (via the decay of heavier charmonium
states such as \Pcgc and \PpsiTwoS), and non-prompt \PJpsi from the decay of long-lived b-hadrons. 
The precise vertexing capabilities of the ALICE detector allow us to determine  
the non-prompt component at mid-rapidity, which 
is discussed in this work.    
This measurement  is combined with results on inclusive \PJpsi production to determine the prompt \PJpsi cross sections, 
which allow a more direct comparison with models describing the charmonium 
production in hadronic interactions 
as compared to 
the inclusive \PJpsi production.  

\section{Data sample and analysis}
The ALICE apparatus~\cite{ALICEjinst,ALICEperf} consists of a central barrel, covering the pseudorapidity region
$|\eta| < 0.9$, a muon spectrometer with $-4 < \eta <-2.5$ coverage,  
and forward and backward detectors employed for triggering, background rejection and event characterisation.  
The central-barrel detectors that have been used to reconstruct $\PJpsi \rightarrow \Pep\Pem$\ decays are the Inner  
Tracking System (ITS) and the Time Projection Chamber (TPC). They are
located inside a large solenoidal magnet with a field strength of 0.5 T.
The ITS~\cite{ITS} consists of six layers of silicon detectors surrounding the
beam pipe at radial positions between 3.9 and 43.0 cm. Its two innermost layers are
composed of Silicon Pixel Detectors (SPD), which provide the spatial resolution to separate
on a statistical basis the prompt and non-prompt \PJpsi components. 
The active volume of the TPC~\cite{TPC} covers the
range along the beam direction $ |z| < 250$~cm relative to the nominal interaction point 
and extends in radial direction from 85 cm to 247 cm. It is the main tracking device in the
central barrel and, in addition, it is used for particle identification via the measurement of the
specific energy loss ($\dd E/\dd x$) in the detector gas.  

This analysis is based on the data sample collected during the 2013 LHC p-Pb run,  
corresponding to an integrated luminosity 
$\lumint = 51.4 \pm 1.9 \; {\rm \mu b}^{-1}$.  
The events were selected using a minimum-bias trigger provided by the V0 detector~\cite{VZERO}, a system of two 
arrays of 32 scintillator tiles each covering the full azimuth within
$2.8<\eta < 5.1$ (V0A) and $-3.7< \eta < -1.7$ (V0C). 
The trigger required  at least one hit in both the V0A and the V0C scintillator
arrays, and 
the non-single-diffractive p-Pb collisions were selected with an efficiency higher than $>99\%$. 
A radiator-quartz detector, the T0 system~\cite{T0}, provided a measurement of the time of the collisions.  
The V0 and T0 time resolutions allowed discrimination of beam--beam interactions from background events in the 
interaction region. Further background suppression was applied in the offline analysis using temporal information 
from the neutron Zero Degree Calorimeters~\cite{ZDC1,ZDC2}.  

The reconstruction of the  \PJpsi in the $\Pep\Pem$ decay channel is described in detail in 
reference~\cite{ALICEinclusive}. 
The tracks were reconstructed with the ITS and TPC detectors and  
required to have  $\Pt > 1.0$~GeV/$c$ and $|\eta|<0.9$, a minimum number of 70 TPC clusters
per track (out of a maximum of 159), a $\chi^2$ per space point of the  
track fit lower than 4, and at least one hit in the SPD.  
Electrons and positrons selection was based on the ${\rm d}E/{\rm d}x$ values measured in the TPC: the 
${\rm d}E/{\rm d}x$ signal was required to be compatible with the mean electron energy loss within $\pm 3 \sigma$, where $\sigma$ denotes the 
resolution of the ${\rm d}E/{\rm d}x$ measurement. Furthermore, tracks consistent with the pion and proton assumptions were rejected. 
Electrons and positrons that, when paired, were found compatible  
with being result of photon conversions 
were also removed, in
order to reduce the combinatorial background. It was verified, using  
a Monte Carlo simulation, that this procedure does not affect the \PJpsi signal. 
\PJpsi candidates were then obtained by pairing the selected positron and electron candidates in the same event and requiring 
the \PJpsi rapidity 
to be within $-1.37 < y < 0.43 $  (i.e. $|y_{\rm lab}|<0.9$ in the laboratory system).     
The condition that at least one of the two decay tracks has a hit  
in the innermost SPD layer was also required in order to enhance the resolution of the \PJpsi decay vertices.  

The measurement of the fraction of the \PJpsi yield originating from b-hadron decays, $f_{\rm b}$,
relies on the discrimination of \PJpsi mesons produced at a distance from the primary  p-Pb collision
vertex. The pseudoproper decay length variable $x$ is defined as $x = c \cdot \vec{L} \cdot \vec{p_{\rm T}} \cdot m_{\PJpsi} / \Pt$, 
where $\vec{L}$ is the vector pointing from the primary vertex to the \PJpsi~decay vertex and $m_{\PJpsi}$ is the \PJpsi pole mass value~\cite{PDG}.  
The $x$\ resolution is about 150~$\mu {\rm m}$ (60~$\mu {\rm m}$) for \PJpsi of $p_{\rm T} = 1.5$~GeV/$c$ (5~GeV/$c$).   
This allows to determine the fraction of \PJpsi from the decay of b-hadrons for events with \PJpsi \pt\ greater than 1.3~GeV/$c$.
The same approach used in similar analyses for the pp~\cite{ALICEpp} and  
Pb-Pb~\cite{ALICEPbPb} colliding systems is adopted here. It is   
based on an unbinned two-dimensional fit, which is performed by 
minimising the opposite of the logarithm 
of the likelihood function $ \mathcal{L}(m_{\Pep\Pem},x$),
\begin{equation}
	- \ln \mathcal{L}(m_{\Pep\Pem},x) = - \sum_{1}^{N}\ln \left[ \fsig 
        \cdot \ffs(x) 
	\cdot M_{\rbt{Sig}}(m_{\Pep\Pem}) + (1 - \fsig) 
        \cdot \ffb(x) 
	\cdot M_{\rbt{Bkg}}(m_{\Pep\Pem}) \right], 
\label{LogL}
\end{equation}
where $N$ is the number of $\Pep\Pem$ pairs in the invariant mass range $2.2 <
m_{\Pep\Pem} < 4.0$~\gevcc,  
  $\ffs(x)$ and $\ffb(x)$ are Probability Density Functions (PrDFs)
  describing the pseudoproper decay length distribution
  for signal (prompt and non-prompt \PJpsi) and background candidates,
  respectively. Similarly, $M_{\rbt{Sig}}(m_{\Pep\Pem})$ and
  $M_{\rbt{Bkg}}(m_{\Pep\Pem})$ are the PrDFs describing the \Pep\Pem invariant 
  mass distributions for the two components. 
  The signal fraction \fsig\ is defined as the ratio
  of the number of signal candidates over the sum of signal and
  background candidates.  
  The fraction of non-prompt \PJpsi enters into $\ffs(x)$ as:
\begin{equation}
	\ffs(x) = \fbprime \cdot F_{\rbt{b}}(x) + (1 - \fbprime) 
	                       \cdot F_{\rbt{prompt}}(x) ,  
\end{equation}
where $F_{\rbt{prompt}}(x)$ and $F_{\rbt{b}}(x)$ are the PrDFs for prompt and
non-prompt \PJpsi, respectively, and \fbprime\ is the uncorrected fraction
of \PJpsi\ coming from b-hadron decays.   
A small correction due
to the different acceptance times efficiency, averaged over \pt\ 
in a given \pt\  interval ($\langle \acceff \rangle $)  
for prompt and non-prompt \PJpsi\ is necessary to obtain \fb\
from \fbprime:
\begin{equation}
	\label{fbCorr}
	  \fb = \left (1 + \frac{1 - \fbprime}{\fbprime} 
         \frac{\langle \acceff \rangle_{\rbt{b}}}
          {\langle \acceff \rangle_{\rbt{prompt}}} \right )^{-1} .
\end{equation}
The difference in  $\langle \acceff \rangle $ originates from the different \pt\ distributions of prompt and non-prompt \PJpsi 
and the assumption on their spin alignment, as discussed later.  
The different components entering into   
the determination of \fb\ are described in detail in~\cite{ALICEpp,ALICEPbPb}. 
An improved 
procedure was introduced in this analysis to determine  
the resolution function, $R(x)$, which  
describes the accuracy by which $x$\ can be reconstructed and is the key ingredient  
of $F_{\rbt{prompt}}(x)$, $F_{\rbt{b}}(x)$  and $\ffb(x)$.  
$R(x)$ was determined using Monte Carlo simulations and considering the $x$ distributions of prompt \PJpsi reconstructed  with the same procedure and selection criteria  as for data.  
It was parameterised with a double-Gaussian core and a power function ($\propto |x|^{-\lambda}$) for the tails~\cite{ALICEpp}.  
A tuning of the Monte Carlo simulation was applied  to minimise the
residual discrepancy between data and simulation 
for the distribution of the impact parameter of single charged tracks.  
The systematic uncertainty related to the incomplete knowledge of $R(x)$ was thus reduced, as discussed later.  

\begin{figure}[tb]
\centering
\includegraphics[width=.49\textwidth]{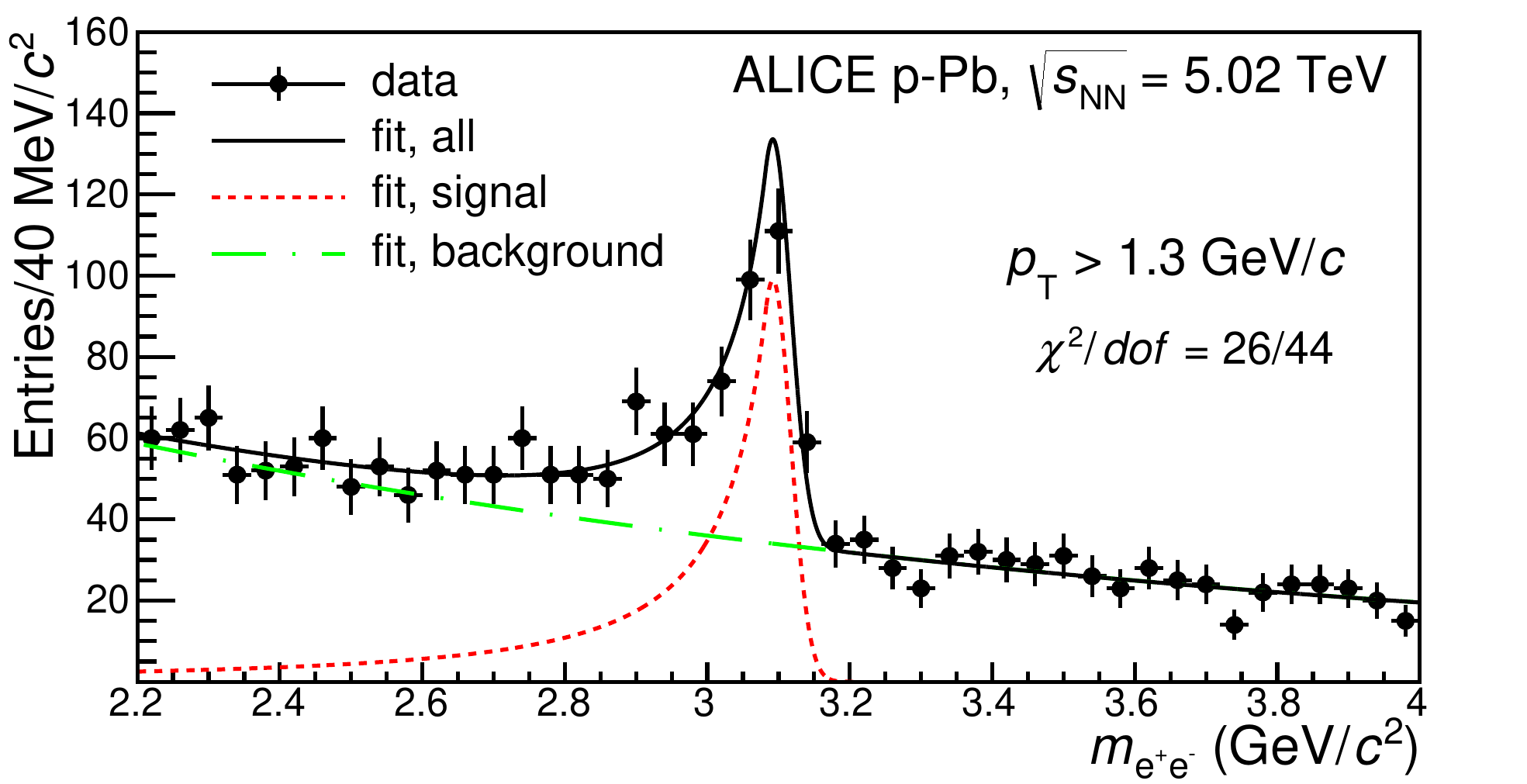}
\includegraphics[width=.49\textwidth]{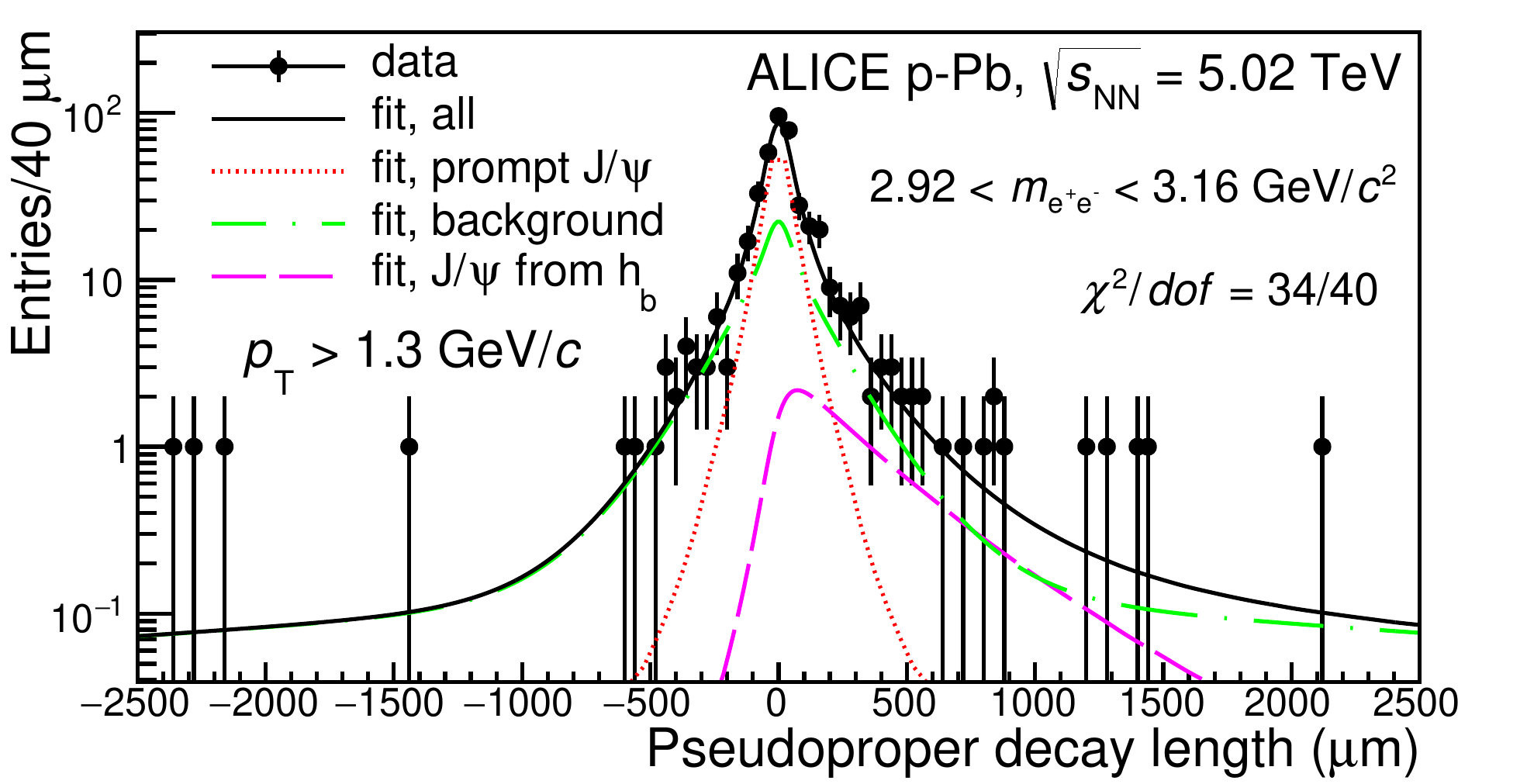}
\caption{Invariant mass (left panel) and  pseudoproper decay length (right panel)
         distributions for \PJpsi candidates with $\Pt >1.3$~GeV/$c$   
         with superimposed  projections  of the maximum likelihood fit. 
	 The latter distribution is limited to the \PJpsi candidates under the
	 mass peak, i.e. for $2.92 < m_{\Pep\Pem} < 3.16 $ GeV/$c^2$, for display purposes only.
         The $\chi^2$ values of these projections are also reported for both distributions.
}
\label{fig:0}
\end{figure}

\begin{figure}[htb]
\centering
\includegraphics[width=.49\textwidth]{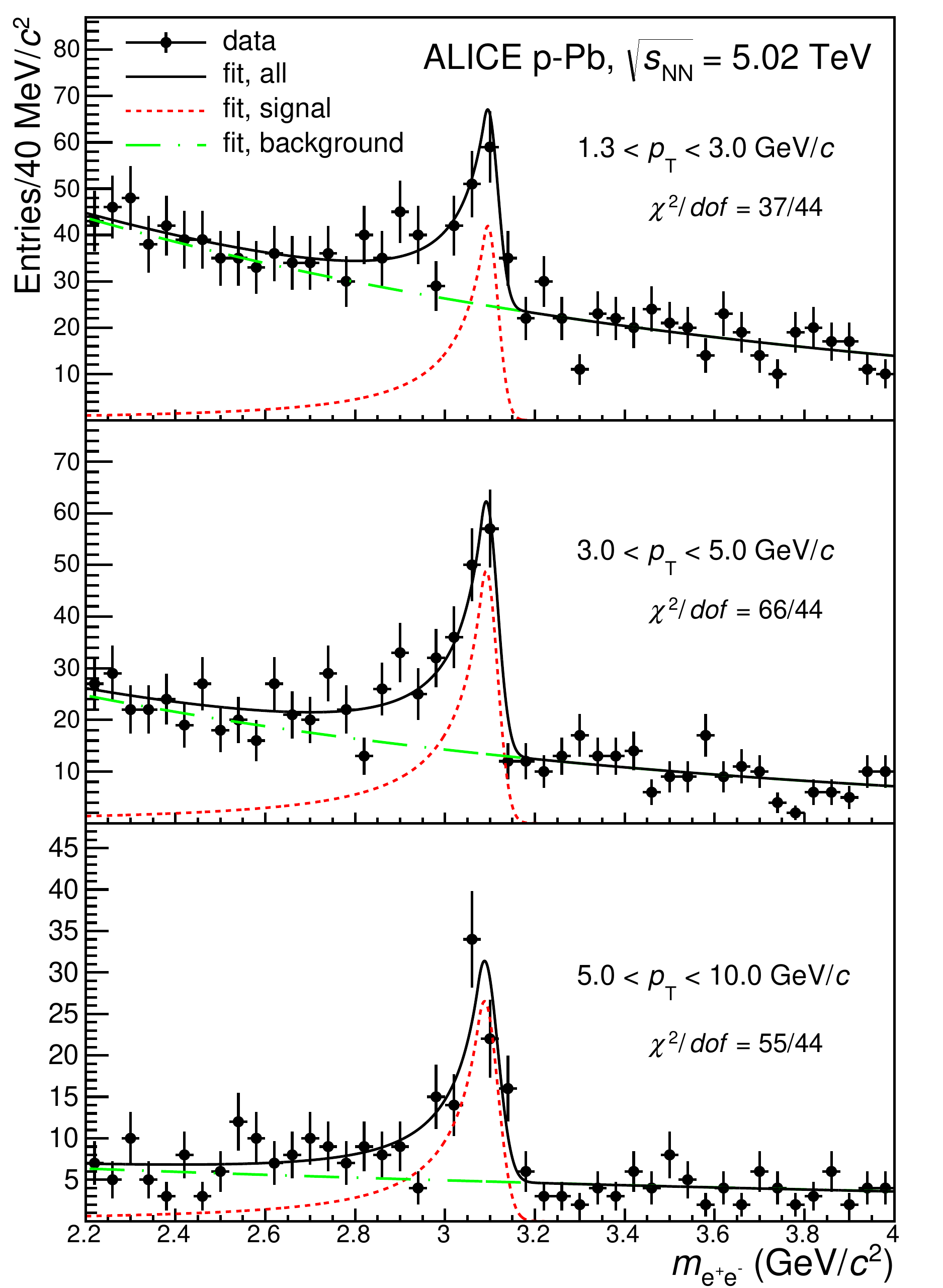}
\includegraphics[width=.49\textwidth]{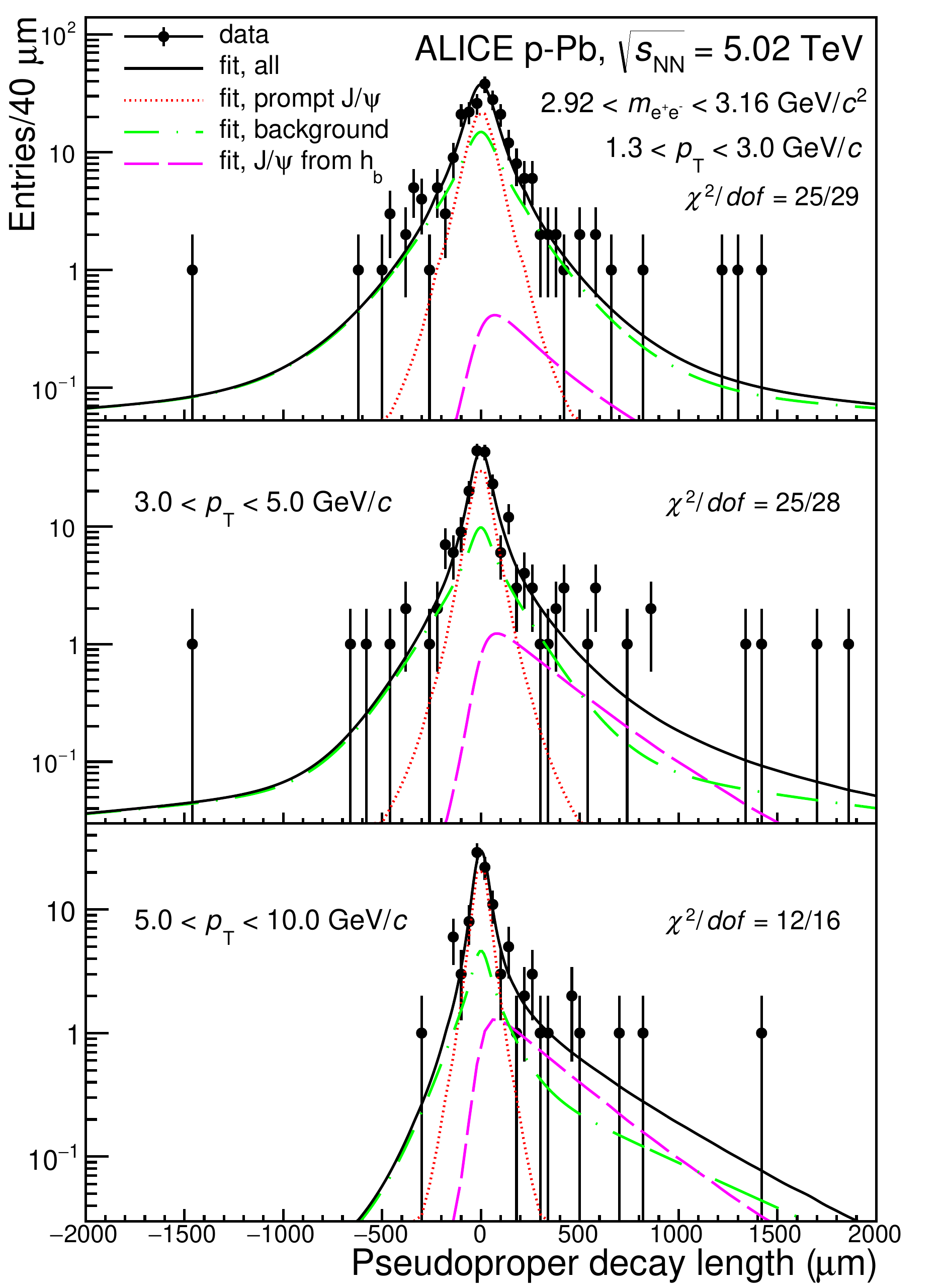}
\caption{Invariant mass (left panels) and  pseudoproper decay length (right panels)
         distributions in different \pt\ intervals  
         with superimposed  projections  of the maximum likelihood fit. 
	 The $x$  distributions are limited to the \PJpsi candidates under the mass peak.
         The $\chi^2$ values of these projections are also reported for all distributions.
}
\label{fig:1}
\end{figure}

In Fig.~\ref{fig:0} the distributions of the invariant mass and the pseudoproper decay length 
for opposite-sign electron pairs with $\Pt > 1.3$~GeV/$c$ are shown with 
superimposed projections of the 
likelihood fit result.  
Although the \PJpsi signal yield is not large, amounting to 360 counts for $\Pt>1.3$~GeV/$c$, the data sample
could be divided into three \pt\ intervals (1.3--3.0, 3.0--5.0 and 5.0--10~GeV/$c$), and the fraction \fb\ was  
evaluated in each interval with the same technique. At low \pt\ there are more candidates, but  
the resolution is worse and the signal over background, $S/B$, is smaller (i.e. $f_{\rm Sig}$ is smaller).
At higher \pt\ the number of candidates is smaller, but the resolution improves and the background becomes
minor.  
In Fig.~\ref{fig:1} the distributions of the invariant mass and of the pseudoproper decay
length are shown in different \pt\ intervals with superimposed projections of the best fit functions.  

The values of the fraction of non-prompt \PJpsi  are evaluated   
with Eq.~\ref{fbCorr} assuming unpolarised prompt \PJpsi. 
The relative variations of \fb\ expected in extreme scenarios   
for the polarisation of prompt \PJpsi  were studied in~\cite{ALICEpp}.
For non-prompt \PJpsi, a small polarisation is obtained using EvtGen~\cite{EVTGEN} as the result of 
the averaging effect caused by the admixture of various exclusive b-hadron decay channels. 
The extreme assumption of a null polarisation also for non-prompt  
\PJpsi  results in a relative decrease of \fb\ by only 1\% at  \pt\ of about 10~GeV/$c$ and ~4\%  
at lower \pt\ (1.3-3.0~GeV/$c$).  
The uncertainties related to the  
polarisation of prompt and non-prompt \PJpsi  are not further propagated to the results, this choice 
being motivated by the small degree of polarisation measured in pp collisions at $\sqrts = 7$~TeV~\cite{LHCBpol,ALICEpol,CMSpol}.

The \pt\ and $y$ distributions used as input to the Monte Carlo simulations 
assume for prompt \PJpsi the shape from next-to-leading order (NLO) Color Evaporation Model (CEM) calculations~\cite{CEMa,CEMb,CEM}, and   
take into account nuclear effects according to the EPS09 parameterisation~\cite{EPS09}. For the non-prompt \PJpsi, 
b-hadrons were generated using PYTHIA 6.4.21~\cite{PYTHIA} with the Perugia-0 tune~\cite{Perugia0} and the 
nuclear shadowing provided by the EPS09 parameterisation was also 
introduced.
In both cases the signal events were injected into p-Pb collisions simulated with HIJING~\cite{HIJING}, and a full simulation of 
the detector response was performed 
adopting GEANT3~\cite{GEANT3} as particle transport code.   
The particle decay was simulated with the EvtGen package~\cite{EVTGEN}, using the PHOTOS model~\cite{PHOTOS}  
to properly describe the \PJpsi radiative decay channel ($\PJpsi\rightarrow\Pep\Pem\gamma$).  
The same reconstruction procedure and selection criteria  were applied to simulated events as to real data.

The systematic uncertainties in the determination of \fb\ arise mainly 
from uncertainties on the  
resolution function, 
and 
the $x$ and $m_{\Pep\Pem}$ PrDFs for  
background pairs, prompt and non-prompt \PJpsi. They were estimated by propagating 
the residual discrepancy between Monte Carlo simulations and data, varying the functional forms assumed for the 
different PrDFs, and repeating the fitting procedure with similar approaches as those described in~\cite{ALICEpp, ALICEPbPb}. 
The uncertainty on the shape of the \pt\ distributions 
in the Monte Carlo simulations 
 introduces also a systematic uncertainty in the determination of \fb. 
 In fact, the Monte Carlo simulations have been used to determine \pt-dependent quantities 
 that were averaged over finite-size \pt\ intervals as, e.g.~$\langle \acceff \rangle$, and the result of 
 the average depends on the \pt\ shape.  
Different assumptions for the \pt\ distributions were considered,  
resulting in variations for the average \pt\ 
of $\sim 15$\%  for both prompt and non-prompt components in the \pt\ integrated sample.  
These include cases without nuclear shadowing,  
a parameterisation of the non-prompt component from perturbative QCD calculations at fixed order with next-to leading-log re-summation
(FONLL)~\cite{FONLL} and a parameterisation of the prompt component with the phenomenological function defined in~\cite{Bossu}.     
Due to the weak \pt\ dependence of \acceff, this uncertainty is found to be significant only for the \pt-integrated 
case~\footnote{A new parameterisation of the nuclear modifications to the PDF, which supersedes EPS09 and has been named as EPPS16,  
has been recently delivered  by the same authors~\cite{EPPS16}. 
Another set of nuclear PDF, nCTEQ15, was also released~\cite{nCTEQ15} and adopted in recent model computations~\cite{JPL}.
The EPS09 parameterisation was used in the Monte Carlo simulation to derive the 
central value of \acceff, but the  alternative assumptions that have been considered produce larger deviations  
in the \pt\ distributions than those obtained using either EPPS16 or nCTEQ15 instead of EPS09.}.  
Table~\ref{tab:systUnc} summarises the systematic uncertainties for the \pt-integrated result ($\Pt>1.3$~GeV/$c$) 
and the three \pt\ intervals.  
%
\begin{table}[tb]
	\centering
	\begin{tabular}{|l|c|c|c|c|} \cline{2-5}
	\multicolumn{1}{c|}{}	 & 
		\multicolumn{4}{c|}{Systematic uncertainties (\%)} \\ \cline{2-5}
	\multicolumn{1}{c|}{}	 & 
	\multicolumn{4}{c|}{ \pt\ range (GeV/$c$)} \\ \cline{1-1}
	Source & \multicolumn{1}{c}{$>1.3$} & \multicolumn{1}{c}{1.3--3} & 
	         \multicolumn{1}{c}{3--5}   & \multicolumn{1}{c|}{5--10} \\
	   \hline
Resolution function                      &  6 & 20 &  4 &  3  \\
PrDF for the $x$ of non-prompt \PJpsi     &  2 &  4 &  1 &  -  \\
PrDF for the $x$ of the background        &  7 & 16 &  6 &  6  \\
MC \pt\ distributions                    &  3 &  1 &  1 &  -  \\
PrDF for the invariant mass of signal     &  6 &  7 &  4 &  3  \\
PrDF for the invariant mass of background &  3 &  8 &  2 &  1  \\
\hline
Total                                    & 12 & 28 &  9 &  7  \\
\hline
\end{tabular}
\caption{
Systematic uncertainties (in percent) on the measurement of the
fraction \fb\ of \PJpsi\ from the decay of b-hadrons, for
different transverse momentum ranges. The symbol ``-'' is used to indicate a 
negligible contribution. 
\label{tab:systUnc}
}
\end{table}

The value of \fb\ in pp collisions at \sqrts~= 5.02~TeV, $\fb^{\rm
  pp}$, is needed to compute the $R_{\rm pPb}$ 
  for prompt and non-prompt \PJpsi~mesons, 
\begin{equation}
\label{eq:rpa-non-prompt}
\begin{split}
	&	  R_{\rm pPb} = \frac{1 - \fb^{\rm pPb}}{1 - \fb^{\rm pp}}
				  \ R_{\rm pPb}^{\rm incl.\ \PJpsi}  \quad \textnormal{for prompt \PJpsi  and} \\
	&  R_{\rm pPb} = \frac{\fb^{\rm pPb}}{\fb^{\rm pp}}
				  \ R_{\rm pPb}^{\rm incl.\ \PJpsi}  \quad \textnormal{for non-prompt \PJpsi,}
\end{split}
\end{equation}
where $R_{\rm pPb}^{\rm incl.\ \PJpsi}$ is the nuclear modification factor for inclusive \PJpsi measured in~\cite{ALICEinclusive}.   
  The same 
  interpolation procedure implemented to derive $\fb^{\rm pp}$ 
  at \sqrts~= 2.76~TeV~\cite{ALICEPbPb} was used to determine $\fb^{\rm pp}$ at \sqrts~= 5.02~TeV. It is based 
  on experimental data (shown in Fig.~\ref{fig:2}) from CDF in $\Pp\Pap$ collisions~\cite{CDF} at lower centre-of-mass energy (1.96 TeV) and from 
  ALICE~\cite{ALICEpp}, ATLAS~\cite{ATLAS7TeV} and CMS~\cite{CMSpp} in pp collisions at higher energy (7 TeV). 
  The value for $\Pt>1.3$~GeV/$c$ is $\fb^{\rm pp} = 0.139 \pm 0.013 $. The values obtained in other \pt\ intervals  
  are reported in the central column of Tab.~\ref{tab:fbpp}.  
\begin{table}[h]
	  \centering
        \begin{tabular}{|l|c|c|} \hline
		\pt\ range (GeV/$c$) &  $\fb^{\rm pp}$ at \sqrts~=5.02 TeV &  $\fb^{\rm pPb}$ at \sqrtsnn~=5.02 TeV \\ \hline
		$> 0 $       &  $0.134 \pm 0.013 $      &  -- \\
		$> 1.3$      &  $0.139 \pm 0.013 $      &  $0.105 \pm 0.038  \pm 0.012 $ \\
		1.3--3 	     &  $0.118 \pm 0.013 $      &  $ < 0.175 $  at 95\%  C.L. \\ 
		3--5	     &  $0.143 \pm 0.012 $      &  $ 0.123 \pm 0.052 \pm 0.011 $  \\
		5--10 	     &  $0.202 \pm 0.013 $      &  $ 0.203 \pm 0.070 \pm 0.014 $  \\ \hline
\end{tabular}
\caption{Fraction of non-prompt \PJpsi in pp collisions at \sqrts~=~5.02~TeV for different \pt\ ranges,  
	as determined with the procedure of interpolation described in~\cite{ALICEPbPb}, and that measured in p-Pb 
	collisions in this analysis. For the latter, the first uncertainty is statistical, the second one is systematical. 
	The upper limit at 95\% confidence level is given for the interval $1.3<\Pt<3$~GeV/$c$.   
\label{tab:fbpp}
}
\end{table}
\section{Results}
\begin{figure}[tb]
\centering{\includegraphics[width=.60\textwidth]{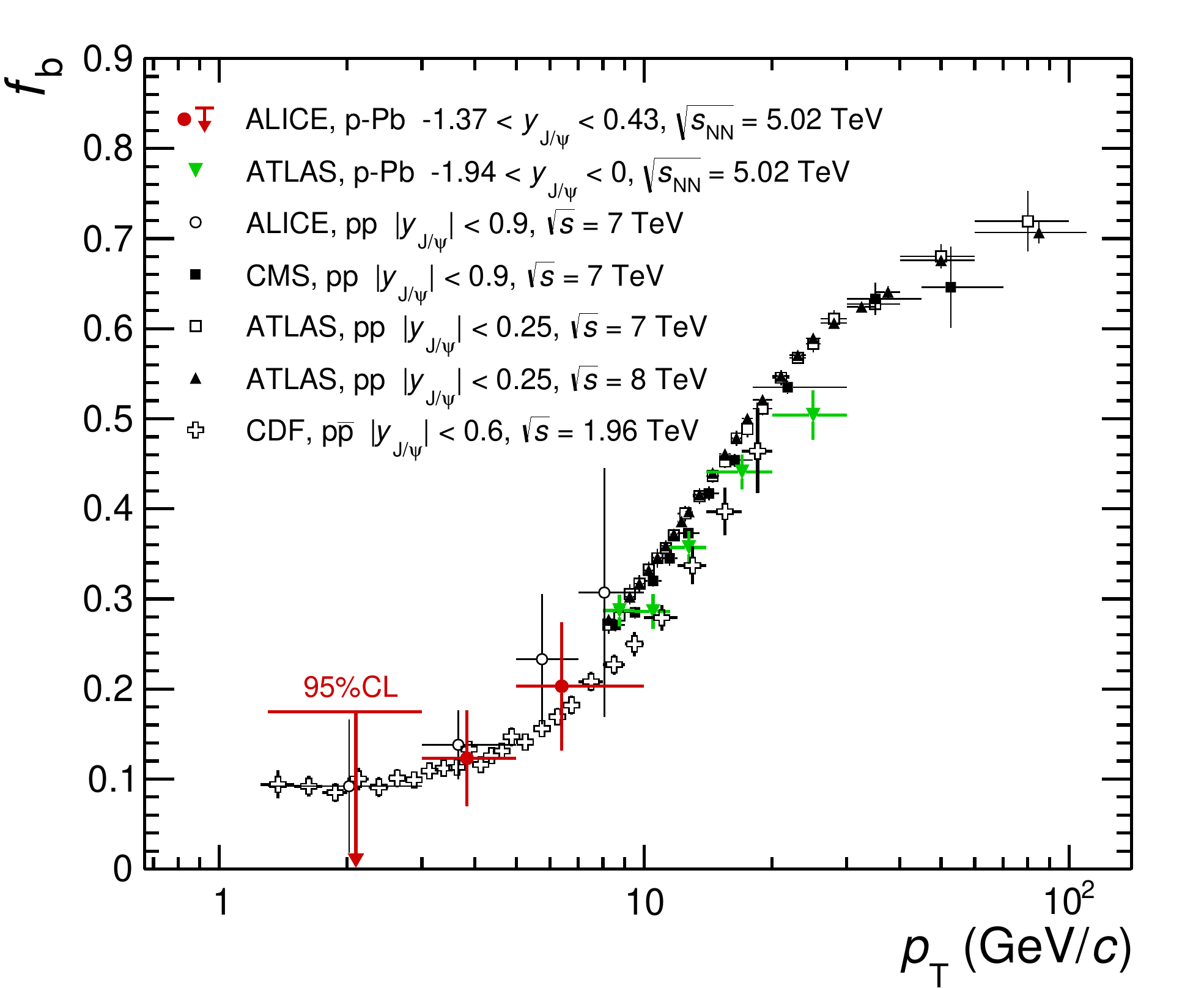}}
\caption{Fraction of \PJpsi from the decay of b-hadrons at mid-rapidity as a function
	 of the \pt\ of \PJpsi in p-Pb collisions at $\sqrt{s_{\rm NN}}=5.02$~TeV compared with results
	 from ATLAS~\cite{ATLAS} in the same colliding system 
	 and results of ALICE~\cite{ALICEpp}, ATLAS~\cite{ATLASpp} and CMS~\cite{CMSpp} 
	 in pp collisions at either $\sqrt{s}=7$~TeV or $\sqrt{s}=8$~TeV.
         Results from CDF~\cite{CDF} in ${\rm p \overline{p}}$ collisions at $\sqrt{s}=1.96$~TeV are also shown. 
	 The ALICE data symbols are placed horizontally at the average value of the \pt\ distribution of each interval (see text for details). 
         For all experiments, the vertical error bars represent the quadratic sum of the statistical and systematic errors.  
	 In the interval $1.3 < \Pt < 3 $~GeV/$c$ the upper limit at the 95\% confidence level is shown, as discussed in the text.  
}
\label{fig:2}
\end{figure}

The fraction of \PJpsi yield originating from decays of b-hadrons in the experimentally accessible kinematic range,  
$p_{\rm t}>1.3$~GeV/$c$ and $-1.37<y<0.43$, which is referred to as ``visible region'' in the following,
is found to be 
\[
	f_{\rm b}=0.105 \pm 0.038\,{\rm (stat.)}  \, \pm 0.012\, {\rm (syst.)}. 
\] 
The results in the different \pt\ intervals are reported in Tab.~\ref{tab:fbpp}. 
In the interval $1.3 < \Pt < 3 $~GeV/$c$ the minimum of Eq.~\ref{LogL}, which is obtained for $\fb=0.05$, is broad and it was not possible 
to define 1$\sigma$ symmetric uncertainty bounds within the physical region $\fb>0$. Therefore an upper limit at the 95\% confidence level 
was derived assuming normally distributed uncertainties.  
Figure~\ref{fig:2} shows the fraction of non-prompt \PJpsi as a function of \pt\ compared to the results of ATLAS~\cite{ATLAS} 
covering the high \pt\ region ($\Pt>8$~GeV/$c$) in a similar rapidity range ($-1.94<y<0$).    
In the figure, the ALICE data symbols are placed horizontally at the average value of the \pt\ distribution of each 
interval. The average was computed using 
the Monte Carlo simulations, which are described in the previous section,  
weighted by the measured \fb.  
In Fig.~\ref{fig:2} the results of  
CDF~\cite{CDF} for ${\rm p \overline{p}}$ collisions at $\sqrt{s}=1.96$~TeV  
and of ALICE~\cite{ALICEpp}, ATLAS~\cite{ATLASpp} and CMS~\cite{CMSpp} experiments 
in pp collisions at either $\sqrt{s}=7$ or $\sqrt{s}=8$~TeV are also shown.  

\begin{figure}[tb]
	\centering{\includegraphics[width=.48\textwidth]{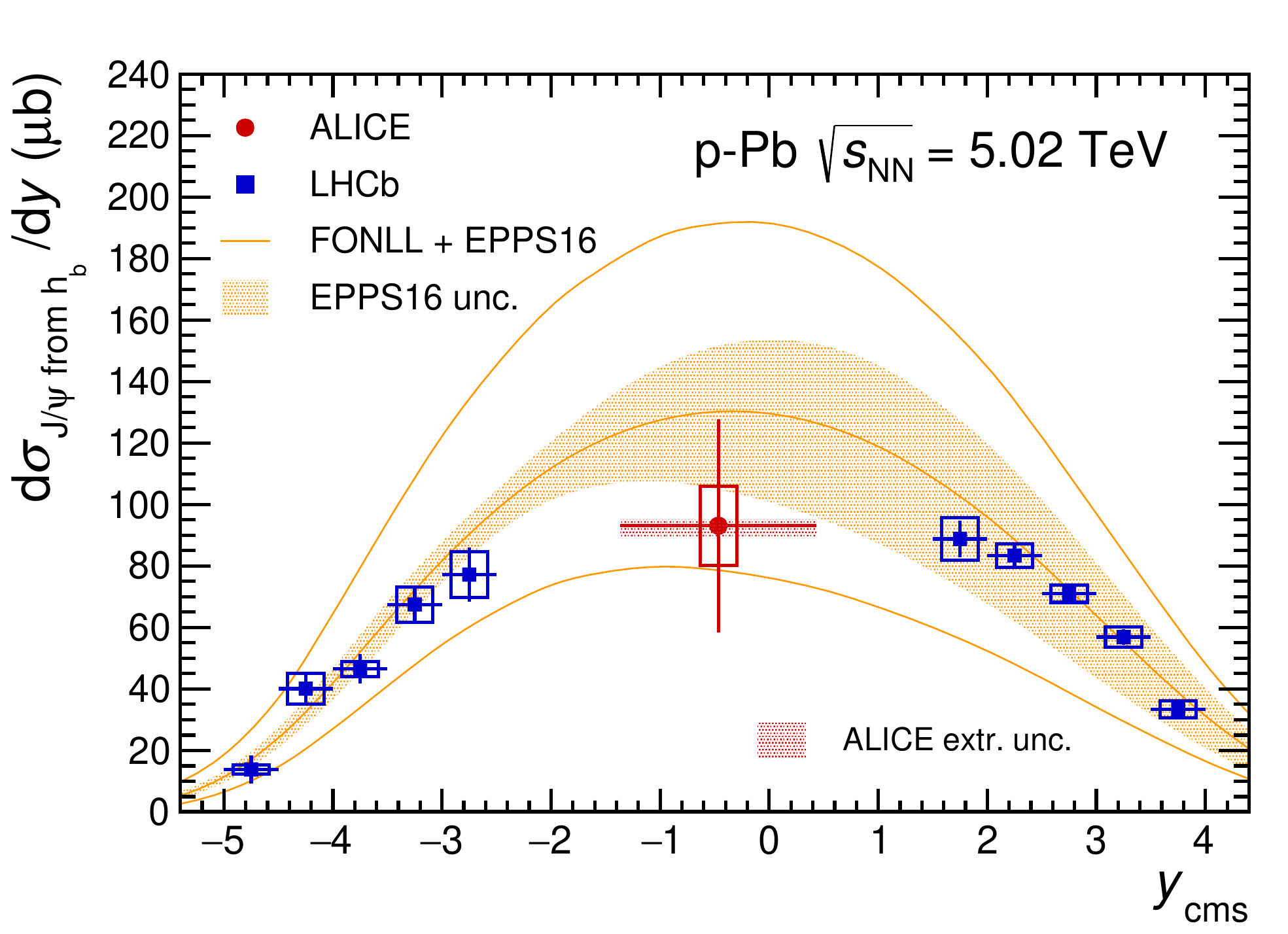}
		   \includegraphics[width=.48\textwidth]{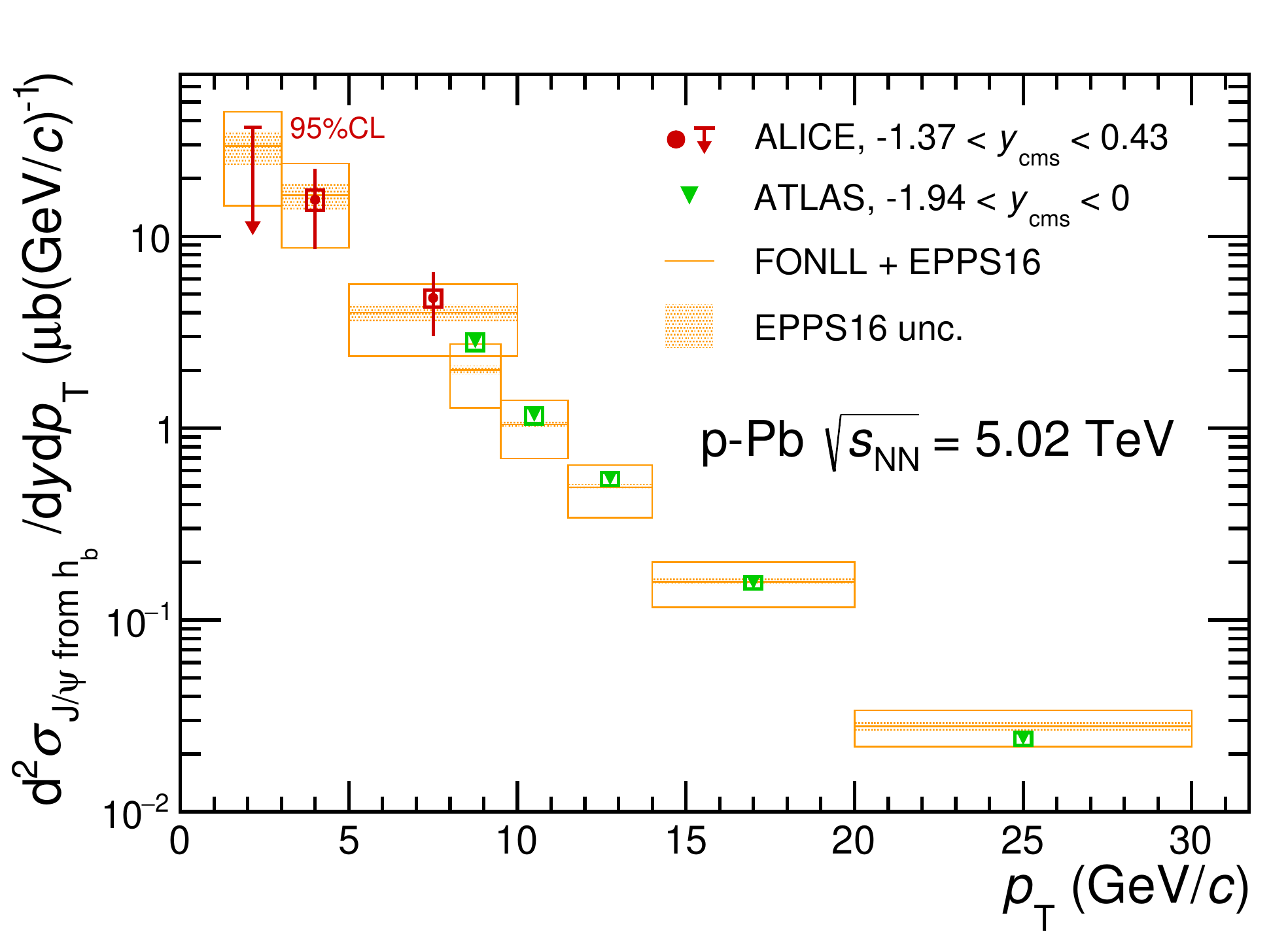}}
	\caption{
		$\dd \sigma_{\rm \PJpsi \, from \, h_b} / \dd y$\ 
		as a function of $y$ (left panel) compared to results obtained in the forward and backward rapidity regions by LHCb~\cite{LHCb} 
		and $\dd^2 \sigma_{\rm \PJpsi \, from \, h_b} / \dd y \, \dd\Pt$\  
		as a function of \pt\ (right panel) compared to ATLAS results~\cite{ATLAS}. 
		The error bars represent the statistical uncertainties,
		while the systematic uncertainties are shown as boxes. 
		In the right panel, the upper limit at the 95\% confidence level is shown with an arrow for the interval $1.3 < \Pt < 3 $~GeV/$c$.  
		The systematic uncertainty on the extrapolation to \pt~=~0 (left panel only) is indicated by the 
		filled red box.  
		Results from FONLL computations~\cite{FONLL} with EPPS16~\cite{EPPS16} nuclear modification 
		of the CTEQ6.6 PDFs~\cite{CTEQ6.6} 
		are shown superimposed, including the    
		total theoretical uncertainty (empty band/boxes) and the EPPS16 contribution (coloured band/boxes).  
		}
	\label{fig:non-prompt}
\end{figure}

By combining the measurement of inclusive \PJpsi cross sections~\cite{ALICEinclusive} with the \fb\ determinations, the prompt and non-prompt 
\PJpsi production cross sections were obtained as follows:
\begin{equation}
	\sigma_{\PJpsi\,{\rm from\, h_b} }=f_{\rm b}\cdot \sigma_{\rm J/\psi} , \quad\quad
	\sigma_{\rm prompt \, J/\psi} = (1-f_{\rm b}) \cdot \sigma_{J/\psi}.
\label{eq:crossSec}
\end{equation}
In the visible region 
the following value is derived for the non-prompt component: 
\[ 
\sigma_{\PJpsi\,{\rm from\, h_b} }^{\rm vis} = 138 \pm 51 {\rm(stat.)} \pm 19 {\rm (syst.)} \; {\rm \mu b}. 
\] 
The visible cross section of non-prompt \PJpsi production was extrapolated down to $\Pt=0$ using FONLL calculations~\cite{FONLL} with CTEQ6.6 
PDFs~\cite{CTEQ6.6} 
and nuclear modification of the parton distribution functions (nPDFs) from the EPPS16 parameterisation~\cite{EPPS16}. The fragmentation of  
b-quarks into hadrons was performed using PYTHIA 6.4.21~\cite{PYTHIA} with the Perugia-0 tune~\cite{Perugia0}.  
The extrapolation factor, which is equal to 
$1.22 ^{+0.02}_{-0.04} $, 
was computed as the ratio of the cross section for $p_{\rm T}^{\rm J/\psi}>0$
and $-1.37<y<0.43$ to 
that in the visible region.   
The uncertainty on the extrapolation factor was determined by combining the FONLL, CTEQ6.6 and EPPS16 uncertainties.  
The FONLL uncertainties have been evaluated 
by varying  the factorisation and renormalisation scales, $\mu_{\rm F}$ and $\mu_{\rm R}$,
independently in the ranges $0.5 < \mu_{\rm F}/m_{\rm T} < 2$, $0.5 < \mu_{\rm R}/m_{\rm T} < 2$,
with the constraint $0.5 < \mu_{\rm F}/\mu_{\rm R} < 2$, where
$m_{\rm T}=\sqrt{p_{\rm T}^2+ m_{\rm b}^2}$. The b-quark mass was varied within
$4.5 <m_{\rm b}< 5.0$~GeV/$c^2$. The CTEQ6.6 and EPPS16 uncertainties  were propagated 
according to the Hessian prescription of the authors of these parameterisations (Eq.~$53$ of reference~\cite{EPPS16}).  
The extrapolated \pt-integrated non-prompt \PJpsi cross section per unit of rapidity is obtained by 
dividing  by the rapidity range $\Delta y=1.8$:   
\[
	\frac{\dd \sigma_{\PJpsi\,{\rm from\, h_b}}}{\dd y} =  
	{\rm 93 \pm 35 \,(stat.)  \pm 13  \, (syst.) ^{+2}_{-3}\, (extr.)} \; {\rm \mu b}.  
\]
In the left panel of Fig.~\ref{fig:non-prompt} this measurement is plotted together with the LHCb~\cite{LHCb} results and compared to 
theoretical predictions based on FONLL pQCD calculations with EPPS16 nPDFs.  
The dashed lines show the total theoretical uncertainties, while the 
coloured band corresponds to the contribution from the EPPS16 uncertainties.  
The cross section was also computed, according to Eq.~\ref{eq:crossSec}, in the three \pt\ intervals and compared to the 
ATLAS measurements~\cite{ATLAS}
for $-1.94<y<0$ and $\Pt>8$~GeV/$c$ (right panel of Fig.~\ref{fig:non-prompt}). 
The ALICE measurement, which covers the low \pt\ region at mid-rapidity, is thus complementary to the data of the other LHC experiments.    
The total theoretical uncertainties on the production cross section, which are dominated by those of the \Pbottom-quark mass and the QCD factorisation and renormalisation scales, are larger 
than the  experimental uncertainties,  
preventing to draw conclusions on the presence of nuclear effects 
for this observable. 

\begin{figure}[tb]
	\centering{\includegraphics[width=.49\textwidth]{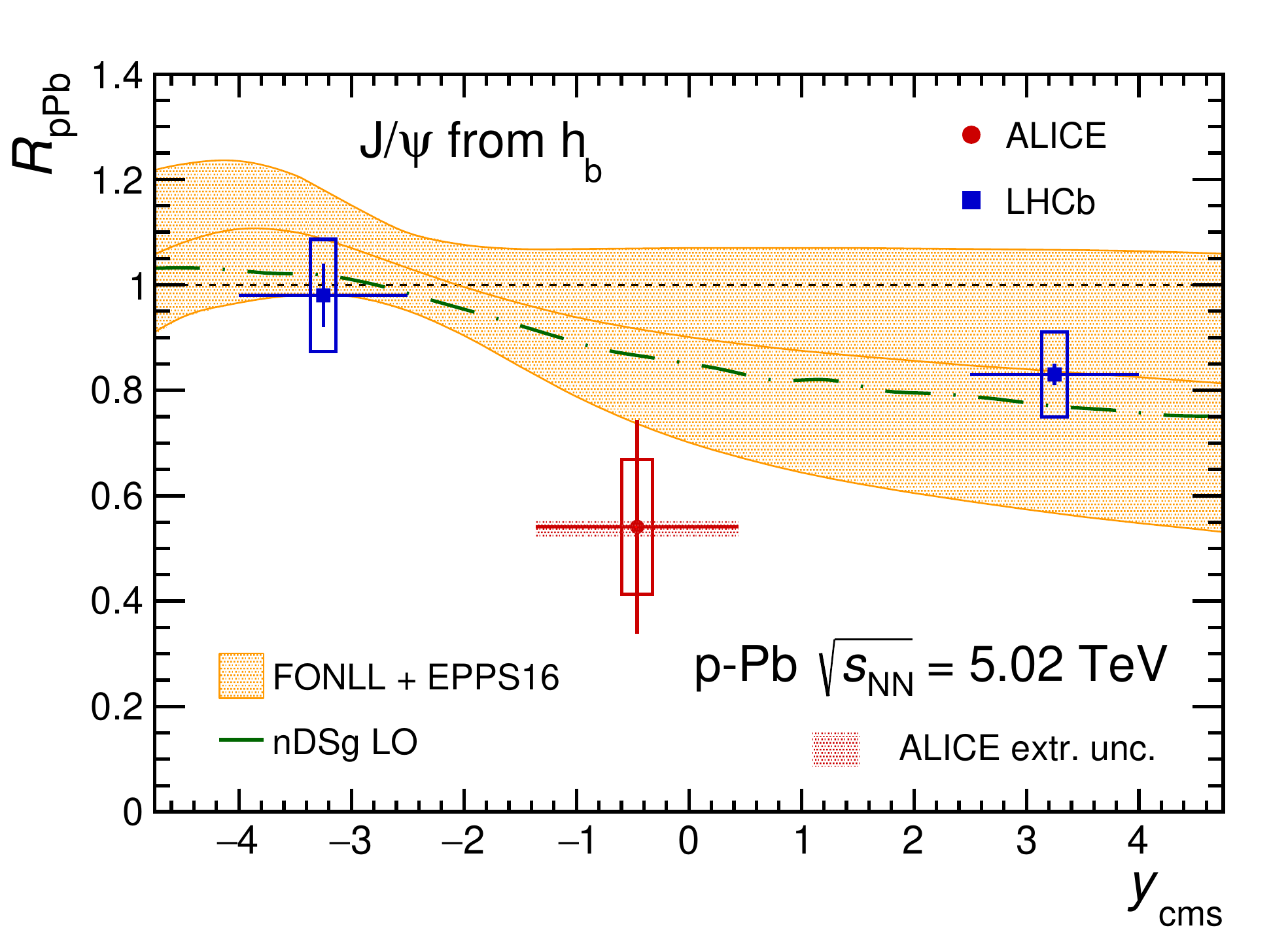}
	           \includegraphics[width=.48\textwidth]{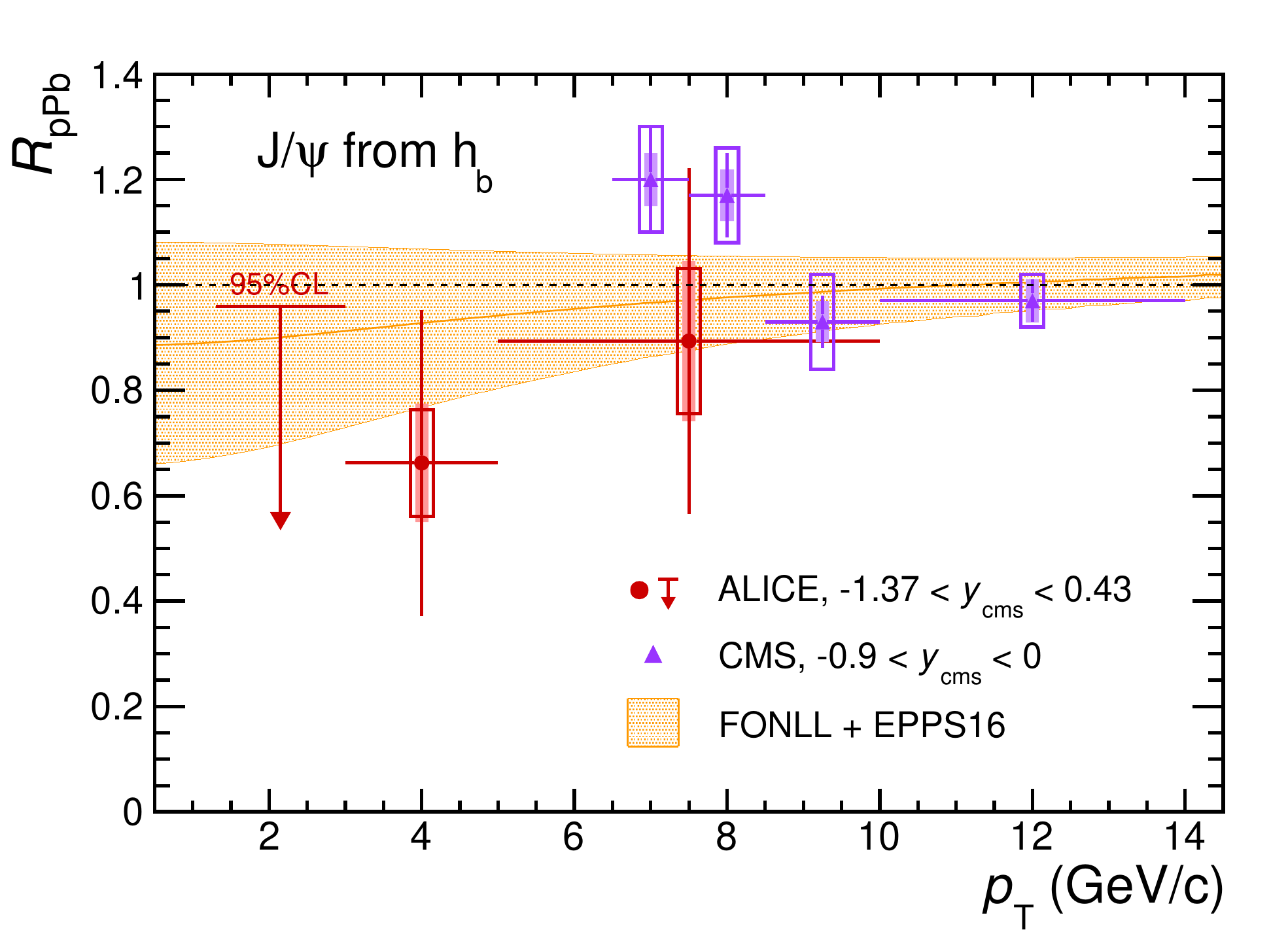}}
	\caption{The nuclear modification factor $R_{\rm pPb}$ of non-prompt \PJpsi as a function of rapidity for $\Pt>0$ (left panel) 
		 and as a function of \pt\ at mid-rapidity (right panel). The error bars and the open boxes indicate, respectively, 
		 the statistical and systematic uncertainties. 
		 In the left hand panel, the results from the LHCb experiment are taken from~\cite{LHCb} 
		 and the systematic uncertainty    
		 on the extrapolation to $\Pt = 0$ for the ALICE data point is depicted by the 
		 filled red box.   
                 In the right hand plot, 
		 the results from the CMS experiment are taken from~\cite{CMS} and the arrow shows the upper 
		 limit at 95\% confidence level for the interval $1.3 < \Pt< 3$~GeV/$c$. 
		 The nuclear modification factors as expected from the EPPS16~\cite{EPPS16} and 
		 the nDSg~\cite{nDSg} (central value shown in the left panel only) 
		 parameterisations are shown superimposed. 
	}
	\label{fig:5}
\end{figure}

The dominant uncertainties of the theoretical predictions cancel out when considering the nuclear 
modification factor $R_{\rm pPb}$, which was determined experimentally according to Eq.~\ref{eq:rpa-non-prompt}.  
Figure~\ref{fig:5} shows the $R_{\rm pPb}$ of non-prompt \PJpsi for $\Pt>0$ as compared to the LHCb measurements at backward and forward 
rapidity~\cite{LHCb} (left panel) and as a function of \pt\ as compared to CMS results~\cite{CMS} (right panel).  
The results are also compared to the FONLL pQCD calculations with EPPS16 nPDFs described previously.    
The central value of an alternative parameterisation of the nuclear PDF, nDSgLO~\cite{nDSg}, 
is also shown for comparison in the left-hand plot.  
The \pt-integrated $R_{\rm pPb}$, which is 
$ R_{\rm pPb}= 0.54 \pm 0.20({\rm stat.)}\, \pm 0.13 ({\rm syst.}) \, ^{+0.01}_{-0.02} ({\rm extr.})$, 
is measured to be 
smaller than unity with a significance of $ 2.3/3.5/1.9 \; \sigma$ (statistical/systematic/combined).
The \pt\ dependence suggests that the suppression of the production originates at low \pt.  

The $\Pbottom\APbottom$ production cross section at mid-rapidity was obtained  
as 
\begin{equation}
	\frac{\dd \sigma_{\Pbottom\APbottom}}{\dd y} 
	  = 
	  \frac{\dd \sigma_{\Pbottom\APbottom}^{\rm model}}{\dd y} \times
	  \frac{  
	  \sigma_{\PJpsi\,{\rm from\, h_b}}^{\rm vis}
		    }{ 
		    \sigma_{\PJpsi\,{\rm from\, h_b}}^{\rm vis, \; model} 
		      } ,
      \label{eq:extr}
\end{equation}
where  
$\dd\sigma_{\Pbeauty\APbeauty}^{\rm model}/\dd y$ and  
$\sigma_{\PJpsi\,{\rm from\, h_b}}^{\rm vis, \; model}$ 
were again obtained  performing FONLL plus  CTEQ6.6 and EPPS16 calculations.   
The average branching fraction of inclusive b-hadron decays to \PJpsi measured at LEP~\cite{Bus92,Adr93,Abr94},
$BR(\rm{h_b\rightarrow J/\psi +X})= (1.16 \pm 0.10)\%$, was used in the computation of 
$\sigma_{\PJpsi\,{\rm from\, h_b}}^{\rm model}$.
The resulting cross section at mid-rapidity is  
\[ \frac{\dd \sigma_{\Pbottom\APbottom}}{\dd y} = 4.1 \pm 1.5 \, ({\rm stat.})  \pm 0.7  ({\rm syst.}) ^{+0.1}_{-0.2} ({\rm extr.}) \; {\rm mb}. \]  

\begin{figure}[tb]
	\centering{
	           \includegraphics[width=.60\textwidth]{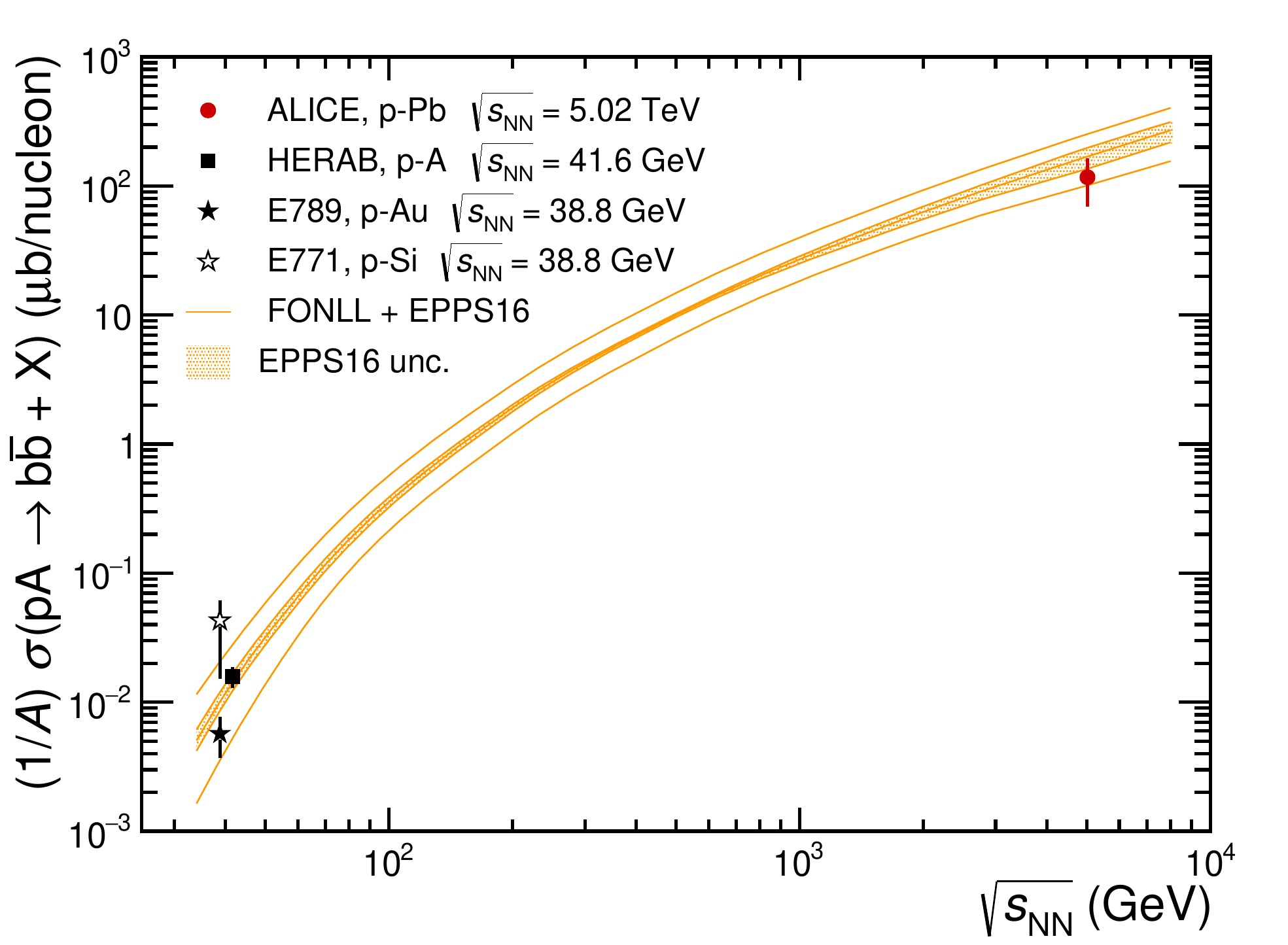}}
	\caption{ Beauty production cross section in p-A collisions as a function of $\sqrt{s_{\rm NN}}$ as 
		  measured by ALICE and at fixed-target experiments (E789~\cite{E789},   
		  E771~\cite{E771} and HERA-B~\cite{HERAB}).  
	          The FONLL calculations with EPPS16 nuclear modification to the PDFs are superimposed in orange.    
		  The full lines show the total theoretical uncertainty, while the coloured band 
		  corresponds to the contribution from the EPPS16 uncertainties. 
	}
	\label{fig:CrossSection}
\end{figure}

The total $\Pbottom\APbottom$ production cross section was computed similarly by extrapolating the visible cross section to the full phase space as  
\begin{equation}
	\sigma({\rm pPb \rightarrow \Pbottom\APbottom}+X) = \alpha_{4\pi} 
	\frac{\sigma_{\PJpsi\,{\rm from\, h_b}}^{vis} 
				     }
				        {2 \cdot \, BR({\rm h_b\rightarrow J/\psi +X})},
\end{equation}
where $\alpha_{4\pi}$ is the ratio between the yield of \PJpsi\ mesons
(from the decay of b-hadrons) in the full phase space and the yield 
in the visible region, 
and the factor 2 takes into account that b-hadrons originate from both \Pbottom and \APbottom quarks.      
The extrapolation factor $\alpha_{4\pi}$ was also 
computed based on FONLL pQCD calculations with EPPS16 nPDFs, with the b-quark fragmentation performed using
PYTHIA 6.4.21 with the Perugia-0 tune,   
and found to be  
$\alpha_{4\pi} = 4.1 \pm 0.2$.      
The resulting cross section is  
\[
 \sigma({\rm p Pb \rightarrow \Pbottom\APbottom}+X) = 
 25 \pm 9 ({\rm stat.}) \pm 4 ({\rm syst.}) \pm 1 ({\rm extr.}) \; {\rm mb} \quad {\rm (ALICE \; only)}.  
 \]
The ALICE measurement is shown in Fig.~\ref{fig:CrossSection} along with the other existing measurements in p-A 
collisions, which were obtained in fixed-target experiments~\cite{E789,E771,HERAB} at lower $\sqrt{s_{\rm NN}}$. 
The experimental results are compared to 
the FONLL calculations using the EPPS16 nPDFs.  

The combination with the LHCb measurements~\cite{LHCb} allows us to extract the total $\Pbottom\APbottom$\ cross section with a significant 
reduction of the uncertainty. The factor $\alpha_{4\pi}$, which is computed as the ratio of the yield in full phase space over 
that covered by ALICE and LHCb, reduces to  $ 1.60^{+0.02}_{-0.03}$ and the total 
cross section becomes  
\[
  \sigma({\rm p Pb \rightarrow \Pbottom\APbottom}+X) = 
 29 \pm 4 ({\rm stat.}) \pm 3 ({\rm syst.}) \pm 1 ({\rm extr.}) \; {\rm mb}  
\quad {\rm (ALICE \; and \; LHCb)}.  
\]

The production cross section of prompt \PJpsi, 
$\dd \sigma_{\rm prompt \, J/\psi} / \dd y$,  
was obtained by 
subtracting the cross section of \PJpsi coming from b-hadron decays 
from the inclusive \PJpsi one measured for $\Pt > 0$~\cite{ALICEinclusive}: 
\[
	\frac{\dd \sigma_{{\rm prompt \, \PJpsi}}}{\dd y} =  
				        {\rm 816 \pm 78 \,(stat.)  \pm 65  \, (syst.) ^{+2}_{-3}\, (extr.)} \; {\rm \mu b}.  
\]
The \pt\ differential cross section was derived using  
Eq.~\ref{eq:crossSec}. The numerical values are reported in Tab.~\ref{tab:Prompt}.  
\begin{table}[ht]
\centering
\resizebox{0.5\columnwidth}{!}{
\begin{tabular}{|c|c|}
\hline
	\pt \, (\gevc)  &  $\dd^2 \sigma^{\rm prompt\,\PJpsi}/\dd y \,  \dd\Pt$  ($\mu$b/(GeV/$c$))   \\[1.01ex]
\hline
1.3~--~3.0  &   200$\pm$35$\pm$25$\pm$8 \\[1.01ex]
3.0~--~5.0    &   111$\pm$15$\pm$8$\pm$4  \\[1.01ex]
5.0~--~10.0 &   18.7$\pm$2.9$\pm$1.2$\pm$0.7  \\[1.01ex]
\hline
\end{tabular}
}
\caption{ The production cross section of prompt \PJpsi\ as a function of \pt\ in p-Pb collisions 
	  at $\sqrtsnn=5.02$~TeV measured for $-1.37<y<0.43$. 
	  The first quoted uncertainty is statistical, the second (third)  is the systematical 
	  one that is correlated (uncorrelated) in \pt.  
\label{tab:Prompt}
}
\end{table}

\begin{figure}[tb]
	\centering{
	           \includegraphics[width=.48\textwidth]{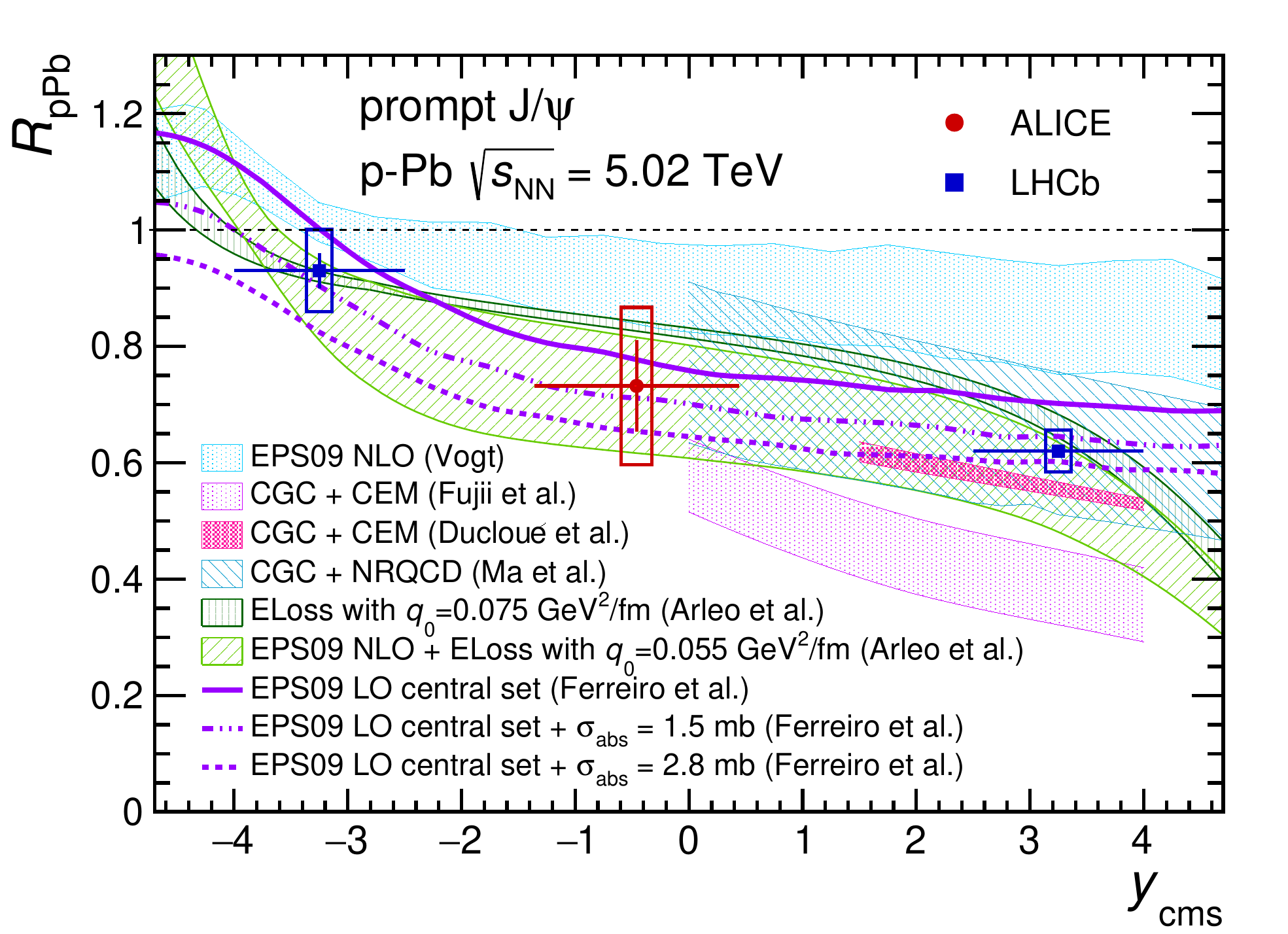}
	           \includegraphics[width=.48\textwidth]{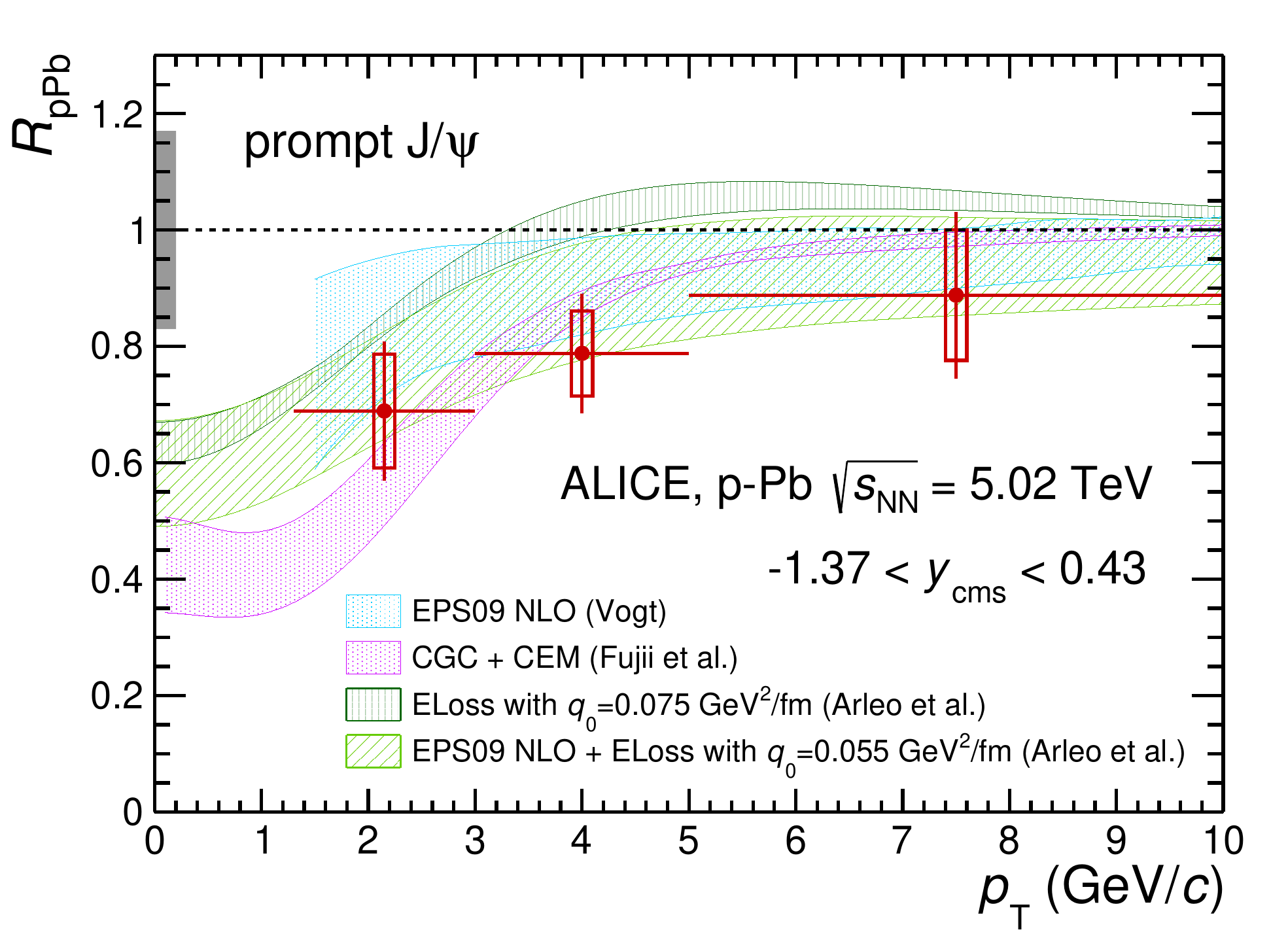}}
	\caption{ 
		$R_{\rm pPb}$ of prompt \PJpsi  versus rapidity (left panel) and 
		  as a function of \pt\ at mid-rapidity (right panel), compared to 
	          theoretical calculations.  
		  Statistical uncertainties are represented by vertical error bars,
		  while open boxes correspond to systematic uncertainties. 
		  Results from the LHCb experiment at backward and forward rapidity 
		  are shown in the left panel~\cite{LHCb}. 
		  The box around $R_{\rm pPb} = 1$ in the right panel 
		  shows the size of the correlated relative uncertainty.
		  Results from various 
		  models~\cite{Vogt,CEM,ARLEO,Ferreiro,Ferreiro2,CGCold,ducloue,ma} 
		  are also shown, see text for details.  
	}
	\label{fig:6}
\end{figure}

The nuclear modification factor for prompt \PJpsi was\ computed using Eq.~\ref{eq:rpa-non-prompt}. 
With respect to the results discussed in~\cite{ALICEinclusive}, where the inclusive \PJpsi production in p-Pb collisions 
at $\sqrt{s_{\rm NN}}=5.02$~TeV was presented, a more direct comparison with model predictions can now be performed. 
Figure~\ref{fig:6} shows the $R_{\rm pPb}$ of prompt \PJpsi compared to predictions from various models. 
The results indicate that the suppression observed at mid-rapidity is a low \pt\ effect, as also argued for non-prompt \PJpsi.  
One calculation (Vogt~\cite{Vogt,CEM}) is based on the NLO CEM for the prompt \PJpsi production and the EPS09 NLO  
shadowing parameterisation. The theoretical uncertainties arise from those in EPS09 
and from the values of the charm quark mass and of the renormalisation and factorisation scales.
A second calculation (Arleo {\em et al.}~\cite{ARLEO}) is  based on a parameterisation of experimental results on prompt \PJpsi  production
in pp collisions, including the effects of coherent energy loss in the cold
nuclear medium with or without introducing shadowing effects according to the EPS09 NLO parameterisation. 
The model of Ferreiro {\em et al.}~\cite{Ferreiro} employs the EPS09 Leading Order (LO) nPDF with or without effects 
from the interaction with a nuclear medium.
The last set of  models are based on different implementations of the CGC effective theory, which assumes a regime 
of gluon saturation (see~\cite{CGC,CGCreview} for reviews), by using either the CEM for the prompt \PJpsi production 
(Fujii {\em et al.}~\cite{CGCold} and Duclou\'e {\em et al.}~\cite{ducloue}) or the non-relativistic QCD (NRQCD) factorisation approach~\cite{NRQCD} (Ma {\em et al.}~\cite{ma}). 
The results suggest the presence of nuclear effects in the low \pt\ region,  but the present uncertainties do not allow us to 
discriminate among the different models.  

\section{Summary}
The production of b-hadrons in p-Pb collisions at $\sqrt{s_{\rm NN}} = 5.02$~TeV through the inclusive decay 
channel ${\rm h_b} \rightarrow \PJpsi + X$ has been measured at mid-rapidity and down to \PJpsi \pt\ of 1.3 GeV/$c$. 
The 
mid-rapidity 
$ \dd \sigma_{\Pbottom\APbottom} / \dd y  $ and the 
total $\Pbottom\APbottom$ cross section, $\sigma_{\Pbottom\APbottom}$,  were derived.   
The nuclear modification factor of beauty production at mid-rapidity, integrated over \pt, is 
$ R_{\rm pPb} = 0.54 \pm 0.20({\rm stat.)}\, \pm 0.13 ({\rm syst.}) \, ^{+0.01}_{-0.02} ({\rm extr.})$
and compatible within uncertainties to expectations  
from the EPPS16 parameterisation of the nuclear modification to the PDFs.    
The production cross section of prompt \PJpsi was obtained by subtracting the non-prompt component  
from a previous measurement of the inclusive \PJpsi production. 
The nuclear modification factor of prompt \PJpsi indicates a reduced production of low \pt\ \PJpsi,   
with respect to expectations from scaled pp collisions,   
but the present uncertainties do not allow us to discriminate among different models.  

%
%

\newenvironment{acknowledgement}{\relax}{\relax}
\begin{acknowledgement}
\section*{Acknowledgements}

The ALICE Collaboration would like to thank all its engineers and technicians for their invaluable contributions to the construction of the experiment and the CERN accelerator teams for the outstanding performance of the LHC complex.
The ALICE Collaboration gratefully acknowledges the resources and support provided by all Grid centres and the Worldwide LHC Computing Grid (WLCG) collaboration.
The ALICE Collaboration acknowledges the following funding agencies for their support in building and running the ALICE detector:
A. I. Alikhanyan National Science Laboratory (Yerevan Physics Institute) Foundation (ANSL), State Committee of Science and World Federation of Scientists (WFS), Armenia;
Austrian Academy of Sciences and Nationalstiftung f\"{u}r Forschung, Technologie und Entwicklung, Austria;
Ministry of Communications and High Technologies, National Nuclear Research Center, Azerbaijan;
Conselho Nacional de Desenvolvimento Cient\'{\i}fico e Tecnol\'{o}gico (CNPq), Universidade Federal do Rio Grande do Sul (UFRGS), Financiadora de Estudos e Projetos (Finep) and Funda\c{c}\~{a}o de Amparo \`{a} Pesquisa do Estado de S\~{a}o Paulo (FAPESP), Brazil;
Ministry of Science \& Technology of China (MSTC), National Natural Science Foundation of China (NSFC) and Ministry of Education of China (MOEC) , China;
Ministry of Science, Education and Sport and Croatian Science Foundation, Croatia;
Ministry of Education, Youth and Sports of the Czech Republic, Czech Republic;
The Danish Council for Independent Research | Natural Sciences, the Carlsberg Foundation and Danish National Research Foundation (DNRF), Denmark;
Helsinki Institute of Physics (HIP), Finland;
Commissariat \`{a} l'Energie Atomique (CEA) and Institut National de Physique Nucl\'{e}aire et de Physique des Particules (IN2P3) and Centre National de la Recherche Scientifique (CNRS), France;
Bundesministerium f\"{u}r Bildung, Wissenschaft, Forschung und Technologie (BMBF) and GSI Helmholtzzentrum f\"{u}r Schwerionenforschung GmbH, Germany;
General Secretariat for Research and Technology, Ministry of Education, Research and Religions, Greece;
National Research, Development and Innovation Office, Hungary;
Department of Atomic Energy Government of India (DAE), Department of Science and Technology, Government of India (DST), University Grants Commission, Government of India (UGC) and Council of Scientific and Industrial Research (CSIR), India;
Indonesian Institute of Science, Indonesia;
Centro Fermi - Museo Storico della Fisica e Centro Studi e Ricerche Enrico Fermi and Istituto Nazionale di Fisica Nucleare (INFN), Italy;
Institute for Innovative Science and Technology , Nagasaki Institute of Applied Science (IIST), Japan Society for the Promotion of Science (JSPS) KAKENHI and Japanese Ministry of Education, Culture, Sports, Science and Technology (MEXT), Japan;
Consejo Nacional de Ciencia (CONACYT) y Tecnolog\'{i}a, through Fondo de Cooperaci\'{o}n Internacional en Ciencia y Tecnolog\'{i}a (FONCICYT) and Direcci\'{o}n General de Asuntos del Personal Academico (DGAPA), Mexico;
Nederlandse Organisatie voor Wetenschappelijk Onderzoek (NWO), Netherlands;
The Research Council of Norway, Norway;
Commission on Science and Technology for Sustainable Development in the South (COMSATS), Pakistan;
Pontificia Universidad Cat\'{o}lica del Per\'{u}, Peru;
Ministry of Science and Higher Education and National Science Centre, Poland;
Korea Institute of Science and Technology Information and National Research Foundation of Korea (NRF), Republic of Korea;
Ministry of Education and Scientific Research, Institute of Atomic Physics and Romanian National Agency for Science, Technology and Innovation, Romania;
Joint Institute for Nuclear Research (JINR), Ministry of Education and Science of the Russian Federation and National Research Centre Kurchatov Institute, Russia;
Ministry of Education, Science, Research and Sport of the Slovak Republic, Slovakia;
National Research Foundation of South Africa, South Africa;
Centro de Aplicaciones Tecnol\'{o}gicas y Desarrollo Nuclear (CEADEN), Cubaenerg\'{\i}a, Cuba and Centro de Investigaciones Energ\'{e}ticas, Medioambientales y Tecnol\'{o}gicas (CIEMAT), Spain;
Swedish Research Council (VR) and Knut \& Alice Wallenberg Foundation (KAW), Sweden;
European Organization for Nuclear Research, Switzerland;
National Science and Technology Development Agency (NSDTA), Suranaree University of Technology (SUT) and Office of the Higher Education Commission under NRU project of Thailand, Thailand;
Turkish Atomic Energy Agency (TAEK), Turkey;
National Academy of  Sciences of Ukraine, Ukraine;
Science and Technology Facilities Council (STFC), United Kingdom;
National Science Foundation of the United States of America (NSF) and United States Department of Energy, Office of Nuclear Physics (DOE NP), United States of America.
\end{acknowledgement}

\bibliographystyle{utphys}   
\bibliography{biblio}

\providecommand{\href}[2]{#2}\begingroup\raggedright\begin{thebibliography}{10}

\bibitem{Factorisation1}
J.~C. Collins, D.~E. Soper, and G.~F. Sterman, ``{Factorization of Hard
  Processes in QCD},'' \href{http://dx.doi.org/10.1142/9789814503266_0001}{{\em
  Adv. Ser. Direct. High Energy Phys.} {\bfseries 5} (1989) 1--91},
  \href{http://arxiv.org/abs/hep-ph/0409313}{{\ttfamily hep-ph/0409313}}.

\bibitem{Factorisation2}
S.~Catani, M.~Ciafaloni, and F.~Hautmann, ``{High-energy factorization and
  small x heavy flavor production},''
  \href{http://dx.doi.org/10.1016/0550-3213(91)90055-3}{{\em Nucl. Phys. B}
  {\bfseries 336} (1991) 135--188}.

\bibitem{Cabibbo}
N.~Cabibbo and G.~Parisi, ``{Exponential hadronic spectrum and quark
  liberation},'' \href{http://dx.doi.org/10.1016/0370-2693(75)90158-6}{{\em
  Phys. Lett. B} {\bfseries 59} (1975) 67}.

\bibitem{Shuryak}
E.~V. Shuryak, ``{Quark-Gluon Plasma and Hadronic Production of Leptons,
  Photons and Pions},''
  \href{http://dx.doi.org/10.1016/0370-2693(78)90370-2}{{\em Phys. Lett. B}
  {\bfseries 150} (1978) 150}.

\bibitem{EMC}
{{\bfseries European Muon} Collaboration}, ``The ratio of the nucleon structure
  functions $f^2_n$ for iron and deuterium,''
  \href{http://dx.doi.org/10.1016/0370-2693(83)90437-9}{{\em Phys. Lett. B}
  {\bfseries 123} (1983) 275}.

\bibitem{shadowing}
N.~Armesto, ``{Nuclear shadowing},''
  \href{http://dx.doi.org/10.1088/0954-3899/32/11/R01}{{\em J. Phys. G.}
  {\bfseries 32} (2016) R367},
  \href{http://arxiv.org/abs/hep-ph/0604108}{{\ttfamily hep-ph/0604108}}.

\bibitem{CGC}
J.-P. Blaizot, F.~Gelis, and R.~Venugopalan, ``{High energy pA collisions in
  the color glass condensate approach II: quark pair production},''
  \href{http://dx.doi.org/10.1016/j.nuclphysa.2004.07.006}{{\em Nucl. Phys. A}
  {\bfseries 743} (2004) 57},
  \href{http://arxiv.org/abs/hep-ph/0402257}{{\ttfamily hep-ph/0402257}}.

\bibitem{CGCreview}
F.~Gelis, ``{Color Glass Condensate and Glasma},''
  \href{http://dx.doi.org/10.1142/S0217751X13300019}{{\em Int. J. Mod. Phys. A}
  {\bfseries 28} (2013) 1330001},
  \href{http://arxiv.org/abs/1211.3327}{{\ttfamily arXiv:1211.3327 [hep-ph]}}.

\bibitem{Vitev}
I.~Vitev, ``{Non-Abelian energy loss in cold nuclear matter},''
  \href{http://dx.doi.org/10.1103/PhysRevC.75.064906}{{\em Phys. Rev. C}
  {\bfseries 75} (2007) 064906},
  \href{http://arxiv.org/abs/hep-th/0703002}{{\ttfamily hep-th/0703002}}.

\bibitem{Lev}
M.~Lev and B.~Petersson, ``{Nuclear effects at large transverse momentum in a
  QCD parton model},'' \href{http://dx.doi.org/10.1007/BF01648792}{{\em Z.
  Phys. C} {\bfseries 21} (1983) 155}.

\bibitem{Wang}
X.~Wang, ``{Systematic study of high $p_{\rm T}$ hadron spectra in pp, pA, and
  AA collisions at ultrarelativistic energies},''
  \href{http://dx.doi.org/10.1103/PhysRevC.61.064910}{{\em Phys. Rev. C}
  {\bfseries 61} (2000) 064910},
  \href{http://arxiv.org/abs/nucl-th/9812021}{{\ttfamily nucl-th/9812021}}.

\bibitem{Kopeliovich}
B.~Kopeliovich, J.~Nemchik, A.~{Sch\"{a}fer}, and A.~Tarasov, ``{Cronin Effect
  in Hadron Production Off Nuclei},''
  \href{http://dx.doi.org/10.1103/PhysRevLett.88.232303}{{\em Phys. Rev. Lett.}
  {\bfseries 88} (2002) 232303},
  \href{http://arxiv.org/abs/hep-ph/0201010}{{\ttfamily hep-ph/0201010}}.

\bibitem{E789}
{{\bfseries E789} Collaboration}, ``{Measurement of the Bottom-Quark Production
  Cross Section in 800 GeV/$c$ Proton-Gold Collisions},''
  \href{http://dx.doi.org/10.1103/PhysRevLett.74.3118}{{\em Phys. Rev. Lett.}
  {\bfseries 74} (1995) 3118}.

\bibitem{E771}
{{\bfseries E771} Collaboration}, ``{Measurement of the $\Pbottom\APbottom$
  Cross Section in 800 GeV/$c$ Proton-Silicon Interactions},''
  \href{http://dx.doi.org/10.1103/PhysRevLett.82.41}{{\em Phys. Rev. Lett.}
  {\bfseries 82} (1999) 41}.

\bibitem{HERAB}
{{\bfseries HERA-B} Collaboration}, ``{Bottom production cross section from
  double muonic decays of b-flavoured hadrons in 920 GeV proton-nucleus
  collisions},'' \href{http://dx.doi.org/10.1016/j.physletb.2007.04.022}{{\em
  Phys. Lett. B.} {\bfseries 650} (2007) 103},
  \href{http://arxiv.org/abs/hep-ex/0612024}{{\ttfamily hep-ex/0612024}}.

\bibitem{PPR2}
{{\bfseries ALICE} Collaboration}, ``{ALICE: Physics Performance Report, Volume
  II},'' \href{http://dx.doi.org/10.1088/0954-3899/32/10/001}{{\em J. Phys.}
  {\bfseries G32} (2006) 1295}.

\bibitem{LHCb}
{{\bfseries LHCb} Collaboration}, ``{Study of \PJpsi production and cold
  nuclear matter effects in pPb collisions at $\sqrtsnn = 5$~TeV},''
  \href{http://dx.doi.org/10.1007/JHEP02(2014)072}{{\em J. High Energ. Phys.}
  {\bfseries 02} (2014) 072}, \href{http://arxiv.org/abs/1308.6729}{{\ttfamily
  arXiv:1308.6729 [nucl-ex]}}.

\bibitem{LHCb8TeV}
{{\bfseries LHCb} Collaboration}, ``{Prompt and nonprompt \PJpsi production and
  nuclear modification in pPb collisions at $\sqrtsnn = 8.16$~TeV},''
  \href{http://dx.doi.org/10.1016/j.physletb.2017.09.058}{{\em Phys. Lett. B}
  {\bfseries 774} (2017) 159},
  \href{http://arxiv.org/abs/1706.07122}{{\ttfamily arXiv:1706.07122
  [hep-ex]}}.

\bibitem{CMSBmeson}
{{\bfseries CMS} Collaboration}, ``{Study of B meson production in p+Pb
  collisions at $\sqrtsnn=5.02$~TeV using exclusive hadronic decays},''
  \href{http://dx.doi.org/10.1103/PhysRevLett.116.032301}{{\em Phys. Rev.
  Lett.} {\bfseries 116} (2016) 032301},
  \href{http://arxiv.org/abs/1508.06678}{{\ttfamily arXiv:1508.06678
  [nucl-ex]}}.

\bibitem{ATLAS}
{{\bfseries ATLAS} Collaboration}, ``{Measurement of differential \PJpsi
  production cross sections and forward-backward ratios in p+Pb collisions with
  the ATLAS detector},''
  \href{http://dx.doi.org/10.1103/PhysRevC.92.034904}{{\em Phys. Rev. C}
  {\bfseries 92} (2015) 034904},
  \href{http://arxiv.org/abs/1505.08141}{{\ttfamily arXiv:1505.08141
  [hep-ex]}}.

\bibitem{CMS}
{{\bfseries CMS} Collaboration}, ``{Measurement of prompt and non-prompt
  $\mathrm{J}/{\psi }$ production in $\mathrm {p}\mathrm {p}$ and $\mathrm
  {p}\mathrm {Pb}$ collisions at $\sqrt{s_{\mathrm {NN}}} =5.02\,\text {TeV}
  $},'' \href{http://dx.doi.org/10.1140%2Fepjc%2Fs10052-017-4828-3}{{\em Eur.
  Phys. J. C} {\bfseries 77} (2017) 269},
  \href{http://arxiv.org/abs/1702.01462}{{\ttfamily arXiv:1702.01462
  [nucl-ex]}}.

\bibitem{ATLAS8TeV}
{{\bfseries ATLAS} Collaboration}, ``{ Measurement of quarkonium production in
  proton--lead and proton--proton collisions at 5.02 TeV with the ATLAS
  detector},'' \href{http://arxiv.org/abs/1709.03089}{{\ttfamily
  arXiv:1709.03089 [nucl-ex]}}.

\bibitem{CMSjet}
{{\bfseries CMS} Collaboration}, ``{Transverse momentum spectra of inclusive b
  jets in p-Pb collisions at $\sqrtsnn = 5.02$~TeV},''
  \href{http://dx.doi.org/10.1016/j.physletb.2016.01.010}{{\em Phys. Lett. B}
  {\bfseries 754} (2016) 59}, \href{http://arxiv.org/abs/1510.03373}{{\ttfamily
  arXiv:1510.03373 [nucl-ex]}}.

\bibitem{ALICEele}
{{\bfseries ALICE} Collaboration}, ``{Measurement of electrons from
  beauty-hadron decays in p-Pb collisions at $\sqrtsnn=5.02$~TeV and Pb--Pb
  collisions at $\sqrtsnn$=2.76~TeV},''
  \href{http://dx.doi.org/10.1007/JHEP07(2017)052}{{\em J. High Energ. Phys.}
  {\bfseries 07} (2017) 52}, \href{http://arxiv.org/abs/1609.03898}{{\ttfamily
  arXiv:1609.03898 [nucl-ex]}}.

\bibitem{ALICEinclusive}
{{\bfseries ALICE} Collaboration}, ``{Rapidity and transverse-momentum
  dependence of the inclusive \PJpsi nuclear modification factor in p-Pb
  collisions at $\sqrtsnn=5.02$~TeV},''
  \href{http://dx.doi.org/10.1007/JHEP06(2015)055}{{\em J. High Energ. Phys.}
  {\bfseries 06} (2015) 055}, \href{http://arxiv.org/abs/1503.07179}{{\ttfamily
  arXiv:1503.07179 [nucl-ex]}}.

\bibitem{BRAMBILLA}
N.~Brambilla {\em et~al.}, ``{Heavy quarkonium: progress, puzzles, and
  opportunities},''
  \href{http://dx.doi.org/10.1140/epjc/s10052-010-1534-9}{{\em Eur. Phys. J. C}
  {\bfseries 71} (2011) 1534}, \href{http://arxiv.org/abs/1010.5827}{{\ttfamily
  arXiv:1010.5827 [hep-ph]}}.

\bibitem{SAPOREGRAVIS}
A.~Andronic {\em et~al.}, ``{Heavy-flavour and quarkonium production in the LHC
  era: from proton-proton to heavy-ion collisions},''
  \href{http://dx.doi.org/10.1140/epjc/s10052-015-3819-5}{{\em Eur. Phys. J. C}
  {\bfseries 76} (2016) 107}, \href{http://arxiv.org/abs/1506.03981}{{\ttfamily
  arXiv:1506.03981 [nucl-ex]}}.

\bibitem{ALICEjinst}
{{\bfseries ALICE} Collaboration}, ``{The ALICE experiment at the CERN LHC},''
  \href{http://dx.doi.org/10.1088/1748-0221/3/08/S08002}{{\em Journal of
  Instrumentation} {\bfseries 3} no.~08, (2008) S08002}.

\bibitem{ALICEperf}
{{\bfseries ALICE} Collaboration}, ``{Performance of the ALICE experiment at
  the CERN LHC},'' \href{http://dx.doi.org/10.1142/S0217751X14300440}{{\em Int.
  J. Mod. Phys. A} no.~29, (2014) 1430044},
  \href{http://arxiv.org/abs/1402.4476}{{\ttfamily arXiv:1402.4476 [nucl-ex]}}.

\bibitem{ITS}
{{\bfseries ALICE} Collaboration}, ``{Alignment of the ALICE Inner Tracking
  System with cosmic-ray tracks},''
  \href{http://dx.doi.org/10.1088/1748-0221/5/03/P03003}{{\em JINST} {\bfseries
  5} (2010) P03003}, \href{http://arxiv.org/abs/1001.0502}{{\ttfamily
  arXiv:1001.0502 [physics.ins-det]}}.

\bibitem{TPC}
J.~Alme {\em et~al.}, ``{The ALICE TPC, a large 3-dimensional tracking device
  with fast readout for ultra-high multiplicity events},''
  \href{http://dx.doi.org/10.1016/j.nima.2010.04.042}{{\em Nucl. Instrum Meth.
  A} {\bfseries 622} (2010) 316}.

\bibitem{VZERO}
{{\bfseries ALICE} Collaboration}, ``{Performance of the ALICE VZERO system},''
  \href{http://dx.doi.org/10.1088/1748-0221/8/10/P10016}{{\em JINST} {\bfseries
  8} (2013) P10016}, \href{http://arxiv.org/abs/1306.3130}{{\ttfamily
  arXiv:1306.3130 [nucl-ex]}}.

\bibitem{T0}
{{\bfseries ALICE} Collaboration}, ``{ALICE Forward Detectors: FMD, T0 and V0:
  Technical Design Report},'' {\em CERN-LHCC-2004-025} .
  \url{https://cds.cern.ch/record/781854}.

\bibitem{ZDC1}
{{\bfseries ALICE} Collaboration}, ``{Measurement of the cross section for
  electromagnetic dissociation with neutron emission in Pb--Pb collisions at
  $\sqrt{s_{\rm NN}} = 2.76$~TeV},''
  \href{http://dx.doi.org/10.1103/PhysRevLett.109.252302}{{\em Phys. Rev. Lett}
  {\bfseries 109} (2012) 252302},
  \href{http://arxiv.org/abs/1203.2436}{{\ttfamily arXiv:1203.2436 [nucl-ex]}}.

\bibitem{ZDC2}
{{\bfseries ALICE} Collaboration}, ``{Centrality dependence of particle
  production in p-Pb collisions at $\sqrt{s_{\rm NN}} = 5.02$~TeV},''
  \href{http://dx.doi.org/10.1103/PhysRevC.91.064905}{{\em Phys. Rev. C}
  {\bfseries 91} (2015) 064905},
  \href{http://arxiv.org/abs/1412.6828}{{\ttfamily arXiv:1412.6828 [nucl-ex]}}.

\bibitem{PDG}
C.~Patrignani and P.~D.~G. others, ``{Review of Particle Physics},''
  \href{http://dx.doi.org/10.1088/1674-1137/40/10/100001}{{\em Chin. Phys. C}
  {\bfseries 40} (2016) 100001}.

\bibitem{ALICEpp}
{{\bfseries ALICE} Collaboration}, ``{Measurement of prompt \PJpsi and beauty
  hadron production cross sections at mid-rapidity in pp collisions at
  $\sqrts=7$~TeV},'' \href{http://dx.doi.org/10.1007/JHEP11(2012)065}{{\em J.
  High Energ. Phys.} {\bfseries 11} (2012) 065},
  \href{http://arxiv.org/abs/1205.5880}{{\ttfamily arXiv:1205.5880 [hep-ex]}}.

\bibitem{ALICEPbPb}
{{\bfseries ALICE} Collaboration}, ``{Inclusive, prompt and non-prompt \PJpsi
  production at mid-rapidity in Pb-Pb collisions at
  $\sqrt{s_{\mathrm{NN}}}=2.76$~TeV},''
  \href{http://dx.doi.org/10.1007/JHEP07(2015)051}{{\em J. High Energ. Phys.}
  {\bfseries 07} (2015) 051}, \href{http://arxiv.org/abs/1504.07151}{{\ttfamily
  arXiv:1504.07151 [nucl-ex]}}.

\bibitem{EVTGEN}
D.~Lange, ``{The EvtGen particle decay simulation package},''
  \href{http://dx.doi.org/10.1016/S0168-9002(01)00089-4}{{\em Nucl. Instrum.
  Meth. A} {\bfseries 462} (2001) 152}.

\bibitem{LHCBpol}
{{\bfseries LHCb} Collaboration}, ``{Measurement of \PJpsi production in pp
  collisions at $\sqrt{s}=7$~TeV},''
  \href{http://dx.doi.org/10.1140/epjc/s10052-011-1645-y}{{\em Eur. Phys. J. C}
  {\bfseries 71} (2011) 1645}, \href{http://arxiv.org/abs/1103.0423}{{\ttfamily
  arXiv:1103.0423 [hep-ex]}}.

\bibitem{ALICEpol}
{{\bfseries ALICE} Collaboration}, ``{\PJpsi polarization in pp collisions at
  $\sqrt{s}=7$~TeV},''
  \href{http://dx.doi.org/10.1103/PhysRevLett.108.082001}{{\em Phys. Rev. Lett}
  {\bfseries 108} (2012) 082001},
  \href{http://arxiv.org/abs/1111.1630}{{\ttfamily arXiv:1111.1630 [hep-ex]}}.

\bibitem{CMSpol}
{{\bfseries CMS} Collaboration}, ``{Measurement of the prompt \PJpsi and \Pgyii
  polarizations in pp collisions at $\sqrt{s}=7$~TeV},''
  \href{http://dx.doi.org/10.1016/j.physletb.2013.10.055}{{\em Phys. Lett. B}
  {\bfseries 727} (2013) 381}, \href{http://arxiv.org/abs/1307.6070}{{\ttfamily
  arXiv:1307.6070 [hep-ex]}}.

\bibitem{CEMa}
H.~Fritzsch, ``{Producing heavy quark flavors in hadronic collisions: a test of
  quantum chromodynamics},''
  \href{http://dx.doi.org/10.1016/0370-2693(77)90108-3}{{\em Phys. Lett. B}
  {\bfseries 67} (1977) 217}.

\bibitem{CEMb}
F.~Halzen, ``{CVC for gluons and hadroproduction of quark flavors},''
  \href{http://dx.doi.org/10.1016/0370-2693(77)90144-7}{{\em Phys. Lett. B}
  {\bfseries 69} (1977) 105}.

\bibitem{CEM}
J.~Albacete {\em et~al.}, ``{Predictions for p+Pb collisions at
  $\sqrtsnn=5$~TeV },'' \href{http://dx.doi.org/10.1142/S0218301313300075}{{\em
  Int. J. Mod. Phys. E} {\bfseries 22} (2013) 1330007},
  \href{http://arxiv.org/abs/1301.3395}{{\ttfamily arXiv:1301.3395 [hep-ph]}}.

\bibitem{EPS09}
K.~Eskola, H.~Paukkunen, and C.~Salgado, ``{EPS09 --- A new generation of NLO
  and LO nuclear parton distribution functions},''
  \href{http://dx.doi.org/10.1088/1126-6708/2009/04/065}{{\em J. High Energ.
  Phys.} {\bfseries 04} (2009) 065},
  \href{http://arxiv.org/abs/0902.4154}{{\ttfamily arXiv:0902.4154 [hep-ph]}}.

\bibitem{PYTHIA}
T.~Sjostrand, S.~Mrenna, and P.~Z. Skands, ``{PYTHIA 6.4 physics and manual},''
  \href{http://dx.doi.org/10.1088/1126-6708/2006/05/026}{{\em J. High Energ.
  Phys.} {\bfseries 05} (2006) 026},
  \href{http://arxiv.org/abs/hep-ph/0603175}{{\ttfamily arXiv:hep-ph/0603175
  [hep-ph]}}.

\bibitem{Perugia0}
P.~Z. Skands, ``{Tuning Monte Carlo generators: The Perugia tunes},''
  \href{http://dx.doi.org/10.1103/PhysRevD.82.074018}{{\em Phys. Rev. D}
  {\bfseries 82} (2010) 074018},
  \href{http://arxiv.org/abs/1005.3457}{{\ttfamily arXiv:1005.3457 [hep-ph]}}.

\bibitem{HIJING}
X.-N. Wang and M.~Gyulassy, ``{Hijing: A Monte Carlo model for multiple jet
  production in pp, pA, and AA collisions},''
  \href{http://dx.doi.org/10.1103/PhysRevD.44.3501}{{\em Phys. Rev. D}
  {\bfseries 44} (1991) 3501}.

\bibitem{GEANT3}
R.~Brun, F.~Carminati, and S.~Giani, ``{GEANT Detector Description and
  Simulation Tool},'' {\em CERN-W-5013} (1994) .

\bibitem{PHOTOS}
E.~Barberio and Z.~Was, ``{PHOTOS: A Universal Monte Carlo for QED radiative
  corrections. Version 2.0},''
  \href{http://dx.doi.org/10.1016/0010-4655(94)90074-4}{{\em Comput. Phys.
  Commun.} {\bfseries 79} (1994) 291}.

\bibitem{FONLL}
M.~Cacciari, S.~Frixione, N.~Houdeau, M.~Mangano, P.~Nason, and G.~Ridolfi,
  ``{Theoretical predictions for charm and bottom production at the LHC},''
  \href{http://dx.doi.org/10.1007/JHEP10(2012)137}{{\em J. High Energ. Phys.}
  {\bfseries 10} (2012) 137}, \href{http://arxiv.org/abs/1205.6344}{{\ttfamily
  arXiv:1205.6344 [hep-ph]}}.

\bibitem{Bossu}
F.~Bossu {\em et~al.}, ``Phenomenological interpolation of the inclusive
  {J}/$\psi$ cross section to proton-proton collisions at {2.76~TeV and
  5.5~TeV},'' \href{http://arxiv.org/abs/1103.2394}{{\ttfamily arXiv:1103.2394
  [nucl-ex]}}.

\bibitem{EPPS16}
K.~Eskola, P.~Paakkinen, H.~Paukkunen, and C.~Salgado, ``{EPPS16: Nuclear
  parton distributions with LHC data},''
  \href{http://dx.doi.org/10.1140/epjc/s10052-017-4725-9}{{\em Eur. Phys. J. C}
  {\bfseries 77} (2017) 163}, \href{http://arxiv.org/abs/1612.05741}{{\ttfamily
  arXiv:1612.05741 [hep-ph]}}.

\bibitem{nCTEQ15}
K.~Kova\v{r}\'{i}k {\em et~al.}, ``{nCTEQ15: Global analysis of nuclear parton
  distributions with uncertainties in the CTEQ framework},''
  \href{http://dx.doi.org/10.1103/PhysRevD.93.085037}{{\em Phys. Rev. D}
  {\bfseries 93} (2016) 085037}.

\bibitem{JPL}
A.~Kusina, J.-P. Lansberg, I.~Schienbein, and H.-S. Shao, ``{Gluon shadowing
  and antishadowing in heavy-flavor production at the LHC},''
  \href{http://arxiv.org/abs/1712.07024}{{\ttfamily arXiv:1712.07024
  [hep-ph]}}.

\bibitem{CDF}
{{\bfseries CDF} Collaboration}, ``{Measurement of the \PJpsi meson and
  b-hadron production cross sections in \Pproton\APproton collisions at $\sqrts
  = 1960$~GeV},'' \href{http://dx.doi.org/10.1103/PhysRevD.71.032001}{{\em
  Phys. Rev. D} {\bfseries 71} (2005) 032001},
  \href{http://arxiv.org/abs/hep-ex/0412071}{{\ttfamily arXiv:hep-ex/0412071
  [hep-ex]}}.

\bibitem{ATLAS7TeV}
{{\bfseries ATLAS} Collaboration}, ``{Measurement of the differential
  cross-sections of inclusive, prompt and non-prompt \PJpsi production in
  proton-proton collisions at $\sqrts=7$~TeV},''
  \href{http://dx.doi.org/10.1016/j.nuclphysb.2011.05.015}{{\em Nucl. Phys. B}
  {\bfseries 850} (2011) 387}, \href{http://arxiv.org/abs/1104.3038}{{\ttfamily
  arXiv:1104.3038 [hep-ex]}}.

\bibitem{CMSpp}
{{\bfseries CMS} Collaboration}, ``{Prompt and non-prompt \PJpsi production in
  pp collisions at $\sqrt{s} = 7$~TeV},''
  \href{http://dx.doi.org/10.1140/epjc/s10052-011-1575-8}{{\em Eur. Phys. J. C}
  {\bfseries 71} (2011) 1575}, \href{http://arxiv.org/abs/1011.4193}{{\ttfamily
  arXiv:1011.4193 [hep-ex]}}.

\bibitem{ATLASpp}
{{\bfseries ATLAS} Collaboration}, ``{Measurement of the differential
  cross-sections of prompt and non-prompt production of \PJpsi and \Pgyii in pp
  collisions at $\sqrt{s} = 7$ and 8 TeV with the ATLAS detector},''
  \href{http://dx.doi.org/10.1140/epjc/s10052-016-4050-8}{{\em Eur. Phys. J. C}
  {\bfseries 76} (2016) 283}, \href{http://arxiv.org/abs/1512.03657}{{\ttfamily
  arXiv:1512.03657 [hep-ex]}}.

\bibitem{CTEQ6.6}
P.~Nadolsky, H.-L. Lai, Q.-H. Cao, J.~Huston, J.~Pumplin, D.~Stump, W.-K. Tung,
  and C.-P. Yuan, ``Implications of {CTEQ} global analysis for collider
  observables,'' \href{http://dx.doi.org/10.1103/PhysRevD.78.013004}{{\em Phys.
  Rev. D} {\bfseries 78} (2008) 013004},
  \href{http://arxiv.org/abs/0802.0007}{{\ttfamily arXiv:0802.0007 [hep-ph]}}.

\bibitem{nDSg}
D.~de~Florian, R.~Sassot, P.~Zurita, and M.~Stratmann, ``{Global analysis of
  nuclear parton distributions},''
  \href{http://dx.doi.org/10.1103/PhysRevD.85.074028}{{\em Phys. Rev. D}
  {\bfseries 85} (2012) 074028},
  \href{http://arxiv.org/abs/1112.6324}{{\ttfamily arXiv:1112.6324 [hep-ph]}}.

\bibitem{Bus92}
{{\bfseries ALEPH} Collaboration}, ``Measurements of mean lifetime and
  branching fractions of b hadrons decaying to {J}/$\psi$,''
  \href{http://dx.doi.org/10.1016/0370-2693(92)91581-S}{{\em Phys. Lett. B}
  {\bfseries 295} (1992) 396}.

\bibitem{Adr93}
{{\bfseries L3} Collaboration}, ``$\chi_c$ production in hadronic {Z} decays,''
  \href{http://dx.doi.org/10.1016/0370-2693(93)91026-J}{{\em Phys. Lett. B}
  {\bfseries 317} (1993) 467}.

\bibitem{Abr94}
{{\bfseries DELPHI} Collaboration}, ``{J}/$\psi$ production in the hadronic
  decays of the {Z},''
  \href{http://dx.doi.org/10.1016/0370-2693(94)01385-3}{{\em Phys. Lett. B}
  {\bfseries 341} (1994) 109}.

\bibitem{Vogt}
R.~Vogt, ``{Cold nuclear matter effects on J/$\psi$\ and $\Upsilon$\ production
  at the LHC},'' \href{http://dx.doi.org/10.1103/PhysRevC.81.044903}{{\em Phys.
  Rev. C} {\bfseries 81} (2010) 044903},
  \href{http://arxiv.org/abs/1003.3497}{{\ttfamily arXiv:1003.3497 [hep-ph]}}.

\bibitem{ARLEO}
F.~Arleo, R.~Kolevatov, S.~Peign´e, and M.~Rustamova, ``Centrality and $p_{\rm
  t}$ dependence of {J}/$\psi$ suppression in proton-nucleus collisions from
  parton energy loss,'' \href{http://dx.doi.org/10.1007/JHEP05(2013)155}{{\em
  J. High Energ. Phys.} {\bfseries 05} (2013) 155},
  \href{http://arxiv.org/abs/1304.0901}{{\ttfamily arXiv:1304.0901 [hep-ph]}}.

\bibitem{Ferreiro}
E.~G. Ferreiro, F.~Fleuret, J.~P. Lansberg, and A.~Rakotozafindrabe, ``{Impact
  of the nuclear modification of the gluon densities on J/$\psi$\ production in
  pPb collisions at $\sqrt{s_{NN}}=5$~TeV},''
  \href{http://dx.doi.org/10.1103/PhysRevC.88.047901}{{\em Phys. Rev. C}
  {\bfseries 88} (2013) 047901},
  \href{http://arxiv.org/abs/1305.4569}{{\ttfamily arXiv:1305.4569 [hep-ph]}}.

\bibitem{CGCold}
H.~Fujii and K.~Watanabe, ``{Heavy quark pair production in high energy pA
  collisions: quarkonium},''
  \href{http://dx.doi.org/10.1016/j.nuclphysa.2013.06.011}{{\em Nucl. Phys. A}
  {\bfseries 915} (2013) 1}, \href{http://arxiv.org/abs/1304.2221}{{\ttfamily
  arXiv:1304.2221 [hep-ph]}}.

\bibitem{ducloue}
B.~Duclou\'e, T.~Lappi, and H.~M\"antysaar, ``{Forward J/$\psi$\ production in
  proton-nucleus collisions at high energy},''
  \href{http://dx.doi.org/10.1103/PhysRevD.91.114005}{{\em Phys. Rev. D}
  {\bfseries 91} (2015) 114005},
  \href{http://arxiv.org/abs/1503.02789}{{\ttfamily arXiv:1503.02789
  [hep-ph]}}.

\bibitem{NRQCD}
G.~T. Bodwin, E.~Braaten, and G.~P. Lepage, ``{Rigorous QCD analysis of
  inclusive annihilation and production of heavy quarkonium},''
  \href{http://dx.doi.org/10.1103/PhysRevD.51.1125}{{\em Phys. Rev. D}
  {\bfseries 51} (1995) 1125},
  \href{http://arxiv.org/abs/hep-ph/9407339}{{\ttfamily arXiv:hep-ph/9407339
  [hep-ph]}}.

\bibitem{ma}
Y.-Q. Ma, R.~Venugopalan, and H.-F. Zhang, ``{J/$\psi$\ production and
  suppression in high-energy proton-nucleus collisions},''
  \href{http://dx.doi.org/10.1103/PhysRevD.92.071901}{{\em Phys. Rev. D}
  {\bfseries 92} (2015) 071901(R)},
  \href{http://arxiv.org/abs/1503.07772}{{\ttfamily arXiv:1503.07772
  [hep-ph]}}.

\bibitem{Ferreiro2}
E.~G. Ferreiro, ``{Excited charmonium suppression in proton-nucleus collisions
  as a consequence of comovers},''
  \href{http://dx.doi.org/10.1016/j.physletb.2015.07.066}{{\em Phys. Lett. B}
  {\bfseries 749} (2015) 98}, \href{http://arxiv.org/abs/1411.0549}{{\ttfamily
  arXiv:1411.0549 [hep-ph]}}.

\end{thebibliography}\endgroup

\newpage
\appendix
\section{The ALICE Collaboration}
\label{app:collab}

\begingroup
\small
\begin{flushleft}
S.~Acharya\Irefn{org139}\And 
F.T.-.~Acosta\Irefn{org22}\And 
D.~Adamov\'{a}\Irefn{org94}\And 
J.~Adolfsson\Irefn{org81}\And 
M.M.~Aggarwal\Irefn{org98}\And 
G.~Aglieri Rinella\Irefn{org36}\And 
M.~Agnello\Irefn{org33}\And 
N.~Agrawal\Irefn{org48}\And 
Z.~Ahammed\Irefn{org139}\And 
S.U.~Ahn\Irefn{org77}\And 
S.~Aiola\Irefn{org144}
% ...
\And
\renewcommand\labelenumi{\textsuperscript{\theenumi}~}

\section*{Affiliation notes}
\renewcommand\theenumi{\roman{enumi}}
\begin{Authlist}
\item \Adef{org*}Deceased
\item \Adef{orgI}Dipartimento DET del Politecnico di Torino, Turin, Italy
\item \Adef{orgII}M.V. Lomonosov Moscow State University, D.V. Skobeltsyn Institute of Nuclear, Physics, Moscow, Russia
\item \Adef{orgIII}Department of Applied Physics, Aligarh Muslim University, Aligarh, India
\item \Adef{orgIV}Institute of Theoretical Physics, University of Wroclaw, Poland
\end{Authlist}

\section*{Collaboration Institutes}
\renewcommand\theenumi{\arabic{enumi}~}
\begin{Authlist}
\item \Idef{org1}A.I. Alikhanyan National Science Laboratory (Yerevan Physics Institute) Foundation, Yerevan, Armenia
\item \Idef{org2}Benem\'{e}rita Universidad Aut\'{o}noma de Puebla, Puebla, Mexico
\item \Idef{org3}Bogolyubov Institute for Theoretical Physics, National Academy of Sciences of Ukraine, Kiev, Ukraine
\item \Idef{org4}Bose Institute, Department of Physics  and Centre for Astroparticle Physics and Space Science (CAPSS), Kolkata, India
\item \Idef{org5}Budker Institute for Nuclear Physics, Novosibirsk, Russia
\item \Idef{org6}California Polytechnic State University, San Luis Obispo, California, United States
\item \Idef{org7}Central China Normal University, Wuhan, China
\item \Idef{org8}Centre de Calcul de l'IN2P3, Villeurbanne, Lyon, France
\item \Idef{org9}Centro de Aplicaciones Tecnol\'{o}gicas y Desarrollo Nuclear (CEADEN), Havana, Cuba
\item \Idef{org10}Centro de Investigaci\'{o}n y de Estudios Avanzados (CINVESTAV), Mexico City and M\'{e}rida, Mexico
\item \Idef{org11}Centro Fermi - Museo Storico della Fisica e Centro Studi e Ricerche ``Enrico Fermi', Rome, Italy
\item \Idef{org12}Chicago State University, Chicago, Illinois, United States
\item \Idef{org13}China Institute of Atomic Energy, Beijing, China
\item \Idef{org14}Chonbuk National University, Jeonju, Republic of Korea
\item \Idef{org15}Comenius University Bratislava, Faculty of Mathematics, Physics and Informatics, Bratislava, Slovakia
\item \Idef{org16}COMSATS Institute of Information Technology (CIIT), Islamabad, Pakistan
\item \Idef{org17}Creighton University, Omaha, Nebraska, United States
\item \Idef{org18}Department of Physics, Aligarh Muslim University, Aligarh, India
\item \Idef{org19}Department of Physics, Ohio State University, Columbus, Ohio, United States
\item \Idef{org20}Department of Physics, Pusan National University, Pusan, Republic of Korea
\item \Idef{org21}Department of Physics, Sejong University, Seoul, Republic of Korea
\item \Idef{org22}Department of Physics, University of California, Berkeley, California, United States
\item \Idef{org23}Department of Physics, University of Oslo, Oslo, Norway
\item \Idef{org24}Department of Physics and Technology, University of Bergen, Bergen, Norway
\item \Idef{org25}Dipartimento di Fisica dell'Universit\`{a} 'La Sapienza' and Sezione INFN, Rome, Italy
\item \Idef{org26}Dipartimento di Fisica dell'Universit\`{a} and Sezione INFN, Cagliari, Italy
\item \Idef{org27}Dipartimento di Fisica dell'Universit\`{a} and Sezione INFN, Trieste, Italy
\item \Idef{org28}Dipartimento di Fisica dell'Universit\`{a} and Sezione INFN, Turin, Italy
\item \Idef{org29}Dipartimento di Fisica e Astronomia dell'Universit\`{a} and Sezione INFN, Bologna, Italy
\item \Idef{org30}Dipartimento di Fisica e Astronomia dell'Universit\`{a} and Sezione INFN, Catania, Italy
\item \Idef{org31}Dipartimento di Fisica e Astronomia dell'Universit\`{a} and Sezione INFN, Padova, Italy
\item \Idef{org32}Dipartimento di Fisica `E.R.~Caianiello' dell'Universit\`{a} and Gruppo Collegato INFN, Salerno, Italy
\item \Idef{org33}Dipartimento DISAT del Politecnico and Sezione INFN, Turin, Italy
\item \Idef{org34}Dipartimento di Scienze e Innovazione Tecnologica dell'Universit\`{a} del Piemonte Orientale and INFN Sezione di Torino, Alessandria, Italy
\item \Idef{org35}Dipartimento Interateneo di Fisica `M.~Merlin' and Sezione INFN, Bari, Italy
\item \Idef{org36}European Organization for Nuclear Research (CERN), Geneva, Switzerland
\item \Idef{org37}Faculty of Engineering and Science, Western Norway University of Applied Sciences, Bergen, Norway
\item \Idef{org38}Faculty of Nuclear Sciences and Physical Engineering, Czech Technical University in Prague, Prague, Czech Republic
\item \Idef{org39}Faculty of Science, P.J.~\v{S}af\'{a}rik University, Ko\v{s}ice, Slovakia
\item \Idef{org40}Frankfurt Institute for Advanced Studies, Johann Wolfgang Goethe-Universit\"{a}t Frankfurt, Frankfurt, Germany
\item \Idef{org41}Gangneung-Wonju National University, Gangneung, Republic of Korea
\item \Idef{org42}Gauhati University, Department of Physics, Guwahati, India
\item \Idef{org43}Helmholtz-Institut f\"{u}r Strahlen- und Kernphysik, Rheinische Friedrich-Wilhelms-Universit\"{a}t Bonn, Bonn, Germany
\item \Idef{org44}Helsinki Institute of Physics (HIP), Helsinki, Finland
\item \Idef{org45}Hiroshima University, Hiroshima, Japan
\item \Idef{org46}Hochschule Worms, Zentrum  f\"{u}r Technologietransfer und Telekommunikation (ZTT), Worms, Germany
\item \Idef{org47}Horia Hulubei National Institute of Physics and Nuclear Engineering, Bucharest, Romania
\item \Idef{org48}Indian Institute of Technology Bombay (IIT), Mumbai, India
\item \Idef{org49}Indian Institute of Technology Indore, Indore, India
\item \Idef{org50}Indonesian Institute of Sciences, Jakarta, Indonesia
\item \Idef{org51}INFN, Laboratori Nazionali di Frascati, Frascati, Italy
\item \Idef{org52}INFN, Sezione di Bari, Bari, Italy
\item \Idef{org53}INFN, Sezione di Bologna, Bologna, Italy
\item \Idef{org54}INFN, Sezione di Cagliari, Cagliari, Italy
\item \Idef{org55}INFN, Sezione di Catania, Catania, Italy
\item \Idef{org56}INFN, Sezione di Padova, Padova, Italy
\item \Idef{org57}INFN, Sezione di Roma, Rome, Italy
\item \Idef{org58}INFN, Sezione di Torino, Turin, Italy
\item \Idef{org59}INFN, Sezione di Trieste, Trieste, Italy
\item \Idef{org60}Inha University, Incheon, Republic of Korea
\item \Idef{org61}Institut de Physique Nucl\'{e}aire d'Orsay (IPNO), Institut National de Physique Nucl\'{e}aire et de Physique des Particules (IN2P3/CNRS), Universit\'{e} de Paris-Sud, Universit\'{e} Paris-Saclay, Orsay, France
\item \Idef{org62}Institute for Nuclear Research, Academy of Sciences, Moscow, Russia
\item \Idef{org63}Institute for Subatomic Physics, Utrecht University/Nikhef, Utrecht, Netherlands
\item \Idef{org64}Institute for Theoretical and Experimental Physics, Moscow, Russia
\item \Idef{org65}Institute of Experimental Physics, Slovak Academy of Sciences, Ko\v{s}ice, Slovakia
\item \Idef{org66}Institute of Physics, Bhubaneswar, India
\item \Idef{org67}Institute of Physics of the Czech Academy of Sciences, Prague, Czech Republic
\item \Idef{org68}Institute of Space Science (ISS), Bucharest, Romania
\item \Idef{org69}Institut f\"{u}r Kernphysik, Johann Wolfgang Goethe-Universit\"{a}t Frankfurt, Frankfurt, Germany
\item \Idef{org70}Instituto de Ciencias Nucleares, Universidad Nacional Aut\'{o}noma de M\'{e}xico, Mexico City, Mexico
\item \Idef{org71}Instituto de F\'{i}sica, Universidade Federal do Rio Grande do Sul (UFRGS), Porto Alegre, Brazil
\item \Idef{org72}Instituto de F\'{\i}sica, Universidad Nacional Aut\'{o}noma de M\'{e}xico, Mexico City, Mexico
\item \Idef{org73}iThemba LABS, National Research Foundation, Somerset West, South Africa
\item \Idef{org74}Johann-Wolfgang-Goethe Universit\"{a}t Frankfurt Institut f\"{u}r Informatik, Fachbereich Informatik und Mathematik, Frankfurt, Germany
\item \Idef{org75}Joint Institute for Nuclear Research (JINR), Dubna, Russia
\item \Idef{org76}Konkuk University, Seoul, Republic of Korea
\item \Idef{org77}Korea Institute of Science and Technology Information, Daejeon, Republic of Korea
\item \Idef{org78}KTO Karatay University, Konya, Turkey
\item \Idef{org79}Laboratoire de Physique Subatomique et de Cosmologie, Universit\'{e} Grenoble-Alpes, CNRS-IN2P3, Grenoble, France
\item \Idef{org80}Lawrence Berkeley National Laboratory, Berkeley, California, United States
\item \Idef{org81}Lund University Department of Physics, Division of Particle Physics, Lund, Sweden
\item \Idef{org82}Nagasaki Institute of Applied Science, Nagasaki, Japan
\item \Idef{org83}Nara Women{'}s University (NWU), Nara, Japan
\item \Idef{org84}National and Kapodistrian University of Athens, School of Science, Department of Physics , Athens, Greece
\item \Idef{org85}National Centre for Nuclear Research, Warsaw, Poland
\item \Idef{org86}National Institute of Science Education and Research, HBNI, Jatni, India
\item \Idef{org87}National Nuclear Research Center, Baku, Azerbaijan
\item \Idef{org88}National Research Centre Kurchatov Institute, Moscow, Russia
\item \Idef{org89}Niels Bohr Institute, University of Copenhagen, Copenhagen, Denmark
\item \Idef{org90}Nikhef, National institute for subatomic physics, Amsterdam, Netherlands
\item \Idef{org91}NRC ¿Kurchatov Institute¿ ¿ IHEP , Protvino, Russia
\item \Idef{org92}NRNU Moscow Engineering Physics Institute, Moscow, Russia
\item \Idef{org93}Nuclear Physics Group, STFC Daresbury Laboratory, Daresbury, United Kingdom
\item \Idef{org94}Nuclear Physics Institute of the Czech Academy of Sciences, \v{R}e\v{z} u Prahy, Czech Republic
\item \Idef{org95}Oak Ridge National Laboratory, Oak Ridge, Tennessee, United States
\item \Idef{org96}Petersburg Nuclear Physics Institute, Gatchina, Russia
\item \Idef{org97}Physics department, Faculty of science, University of Zagreb, Zagreb, Croatia
\item \Idef{org98}Physics Department, Panjab University, Chandigarh, India
\item \Idef{org99}Physics Department, University of Jammu, Jammu, India
\item \Idef{org100}Physics Department, University of Rajasthan, Jaipur, India
\item \Idef{org101}Physikalisches Institut, Eberhard-Karls-Universit\"{a}t T\"{u}bingen, T\"{u}bingen, Germany
\item \Idef{org102}Physikalisches Institut, Ruprecht-Karls-Universit\"{a}t Heidelberg, Heidelberg, Germany
\item \Idef{org103}Physik Department, Technische Universit\"{a}t M\"{u}nchen, Munich, Germany
\item \Idef{org104}Research Division and ExtreMe Matter Institute EMMI, GSI Helmholtzzentrum f\"ur Schwerionenforschung GmbH, Darmstadt, Germany
\item \Idef{org105}Rudjer Bo\v{s}kovi\'{c} Institute, Zagreb, Croatia
\item \Idef{org106}Russian Federal Nuclear Center (VNIIEF), Sarov, Russia
\item \Idef{org107}Saha Institute of Nuclear Physics, Kolkata, India
\item \Idef{org108}School of Physics and Astronomy, University of Birmingham, Birmingham, United Kingdom
\item \Idef{org109}Secci\'{o}n F\'{\i}sica, Departamento de Ciencias, Pontificia Universidad Cat\'{o}lica del Per\'{u}, Lima, Peru
\item \Idef{org110}Shanghai Institute of Applied Physics, Shanghai, China
\item \Idef{org111}Stefan Meyer Institut f\"{u}r Subatomare Physik (SMI), Vienna, Austria
\item \Idef{org112}SUBATECH, IMT Atlantique, Universit\'{e} de Nantes, CNRS-IN2P3, Nantes, France
\item \Idef{org113}Suranaree University of Technology, Nakhon Ratchasima, Thailand
\item \Idef{org114}Technical University of Ko\v{s}ice, Ko\v{s}ice, Slovakia
\item \Idef{org115}Technische Universit\"{a}t M\"{u}nchen, Excellence Cluster 'Universe', Munich, Germany
\item \Idef{org116}The Henryk Niewodniczanski Institute of Nuclear Physics, Polish Academy of Sciences, Cracow, Poland
\item \Idef{org117}The University of Texas at Austin, Austin, Texas, United States
\item \Idef{org118}Universidad Aut\'{o}noma de Sinaloa, Culiac\'{a}n, Mexico
\item \Idef{org119}Universidade de S\~{a}o Paulo (USP), S\~{a}o Paulo, Brazil
\item \Idef{org120}Universidade Estadual de Campinas (UNICAMP), Campinas, Brazil
\item \Idef{org121}Universidade Federal do ABC, Santo Andre, Brazil
\item \Idef{org122}University College of Southeast Norway, Tonsberg, Norway
\item \Idef{org123}University of Cape Town, Cape Town, South Africa
\item \Idef{org124}University of Houston, Houston, Texas, United States
\item \Idef{org125}University of Jyv\"{a}skyl\"{a}, Jyv\"{a}skyl\"{a}, Finland
\item \Idef{org126}University of Liverpool, Liverpool, United Kingdom
\item \Idef{org127}University of Split, Faculty of Electrical Engineering, Mechanical Engineering and Naval Architecture, Split, Croatia
\item \Idef{org128}University of Tennessee, Knoxville, Tennessee, United States
\item \Idef{org129}University of the Witwatersrand, Johannesburg, South Africa
\item \Idef{org130}University of Tokyo, Tokyo, Japan
\item \Idef{org131}University of Tsukuba, Tsukuba, Japan
\item \Idef{org132}Universit\'{e} Clermont Auvergne, CNRS/IN2P3, LPC, Clermont-Ferrand, France
\item \Idef{org133}Universit\'{e} de Lyon, Universit\'{e} Lyon 1, CNRS/IN2P3, IPN-Lyon, Villeurbanne, Lyon, France
\item \Idef{org134}Universit\'{e} de Strasbourg, CNRS, IPHC UMR 7178, F-67000 Strasbourg, France, Strasbourg, France
\item \Idef{org135} Universit\'{e} Paris-Saclay Centre d¿\'Etudes de Saclay (CEA), IRFU, Department de Physique Nucl\'{e}aire (DPhN), Saclay, France
\item \Idef{org136}Universit\`{a} degli Studi di Pavia, Pavia, Italy
\item \Idef{org137}Universit\`{a} di Brescia, Brescia, Italy
\item \Idef{org138}V.~Fock Institute for Physics, St. Petersburg State University, St. Petersburg, Russia
\item \Idef{org139}Variable Energy Cyclotron Centre, Kolkata, India
\item \Idef{org140}Warsaw University of Technology, Warsaw, Poland
\item \Idef{org141}Wayne State University, Detroit, Michigan, United States
\item \Idef{org142}Westf\"{a}lische Wilhelms-Universit\"{a}t M\"{u}nster, Institut f\"{u}r Kernphysik, M\"{u}nster, Germany
\item \Idef{org143}Wigner Research Centre for Physics, Hungarian Academy of Sciences, Budapest, Hungary
\item \Idef{org144}Yale University, New Haven, Connecticut, United States
\item \Idef{org145}Yonsei University, Seoul, Republic of Korea
\end{Authlist}
\endgroup
\end{document}